\def\C{\mathbb C}
\def\R{\mathbb R}
\def\Z{\mathbb Z}

\def\A{\mathcal A}
\def\F{\mathcal F}
\def\G{\mathcal G}
\def\H{\mathcal H}
\def\K{\mathcal K}
\def\I{\mathcal I}
\def\J{\mathcal J}

\documentclass[reqno]{amsart}
\usepackage{amssymb}
\usepackage{verbatim}
\usepackage{float}

\newtheorem{thm}{Theorem}

\newtheorem{conj}[thm]{Conjecture}

\def\tr{\operatorname{tr}}
\def\diag{\operatorname{diag}}

\begin{document}

\title{Fine group gradings of the real forms of $sl(4,\C)$, $sp(4,\C)$, and $o(4,\C)$}

\author{Ji\v{r}\'\i\  Patera}
\address{Centre de recherches math\'ematiques,
         Universit\'e de Montr\'eal,
         C.P.6128-Centre ville,
         Montr\'eal, H3C\,3J7, Qu\'ebec, Canada}
\email{patera@crm.umontreal.ca}

\author{Edita Pelantov\'a}
\address{Doppler Institute and  Department of mathematics,
         Faculty of Nuclear Science and Physical
         Engineering,  Czech Technical University,
         Trojanova 13, Praha 2, 120 00,
         Czech Republic}
\email{pelantova@km1.fjfi.cvut.cz}

\author{Milena Svobodov\'a}
\address{Doppler Institute and Department of mathematics,
         Faculty of Nuclear Science and Physical
         Engineering,  Czech Technical University,
         Trojanova 13, Praha 2, 120 00,
         Czech Republic}
\email{svobodov@km1.fjfi.cvut.cz}

\date{\today}
\begin{abstract}
We present an explicit description of the 'fine group gradings'
(i.e. group gradings which cannot be further refined) of the real
forms of the semisimple Lie algebras $sl(4,\C)$, $sp(4,\C)$, and
$o(4,\C)$. All together 12 real Lie algebras are considered, and
the total of 44 of their fine group gradings are listed.

The inclusions $sl(4,\C)\supset sp(4,\C)\supset o(4,\C)$ are an
important tool in our presentation. Systematic use is made of the
faithful representations of the three Lie algebras by $4\times 4$
matrices.

\end{abstract}
\maketitle

\section{Introduction}


Wide diversity of applications of low dimensional semisimple Lie
algebras, real or complex, is witnessed by the innumerable papers
found in the literature, where such Lie algebras play a role. In
the article we consider the real forms of the three semisimple Lie
algebras which are faithfully represented by $4\times 4$ matrices.
More precisely, we have the Lie algebras and the inclusions among
them:
$$
sl(4,\C)\supset sp(4,\C)\supset o(4,\C).
$$
Here $sl(4,\C)$ is the Lie algebra of all traceless matrices
$\C^{4\times4}$. Then $sp(4,\C)$ is the Lie algebra of all
(symplectic) transformations which preserve a skew-symmetric
bilinear form in the 4-space $\C^4$. Finally, $o(4,\C)$ is the Lie
algebra of all (orthogonal) transformations which preserve a
symmetric bilinear form in $\C^4$. The second inclusion is
somewhat misleading and therefore deserves a comment. Due to the
fact that $o(4,\C)\simeq sl(2,\C)\times sl(2,\C)$ is not simple,
one has a choice considering $o(4,\C)$ as linear transformations
in $\C^4$. One can introduce symmetric or skew-symmetric bilinear
form in  $\C^4$ invariant under the matrices of $o(4,\C)$. In the
symmetric case, $o(4,\C)$ contains all such transformations, while
all symplectic transformations are in $sp(4,\C)$ and only some of
them in $o(4,\C)$.

The list of the real forms considered in this paper is the
following:
\begin{align}
sl(4,\C)\ &:\qquad sl(4,\R),\ su^{*}(4),\ su(4,0),\ su(3,1),\ su(2,2)\\
sp(4,\C)\ &:\qquad sp(4,\R),\ usp(4,0),\ usp(2,2)\\
o(4,\C)\ &:\qquad so^{*}(4),\ so(4,0),\ so(3,1),\ so(2,2)
\end{align}

The subject of our paper are the gradings of these real forms. For
a general motivation for studying the gradings, we can point out
\cite{Klimyk1,Klimyk2, P3, P1, StovTol, Weyl}. Applications of the
real forms are too numerous to be mentioned here. For example, let
us just mention that among the orthogonal realizations of $B_2$
there are the Lie algebras of the de~Sitter groups, that some real
forms of the symplectic realization have been intensively applied
in the recent years in nuclear physics, and also in quantum optics
\cite{Rowe1,Rowe2,Rowe3,Sanders}.

A grading carries basic structural information about its Lie
algebra. That is particularly true about the fine grading. One
type of a grading  is obtained by decomposing the Lie algebra into
eigensubspaces of mutually commuting automorphisms. This grading
has an interesting property: Its grading subspaces can be indexed
by elements of an Abelian group. Such gradings are called group
gradings. The most famous example of a grading of this type,
obtained by means of automorphisms from the MAD-group which is the
maximal torus of the corresponding Lie group, is called the root
decomposition. It underlies most of the representation theory. As
was recently shown in \cite{Eld}, not all gradings can be
associated with groups of automorphisms. In this article we only
concentrate on group gradings.

The coarsest non-trivial grading is by the group $\Z_2$ of order
2. There is a one-to-one correspondence between real forms of a
Lie algebra, say $L$, and non-equivalent $\Z_2$ groups inside of
the group of automorphisms of $L$.

Motivation of a physicist to study fine gradings may come from
rather different sources. Let us mention just two: (i) For
physical system which display a symmetry of $L$, it is interesting
to know all MAD-groups, because their eigenvalues are the additive
quantum numbers characterizing the state of such system. (ii) A
meaningful relation between physical systems that display
different symmetry Lie algebras, can be established if the Lie
algebras are related by a process of singular deformations
('contractions'). Grading preserving contractions is a powerful
method for describing the wealth of possible deformations.

A classification of MAD-groups of classical Lie algebras over $\C$
is found in \cite {OLG2}; similarly one finds in \cite{OLG3} the
classification of MAD-groups of their real forms. Practically,
however, concise presentation in \cite{OLG2,OLG3} leaves a
laborious task to a reader interested in specific realization of
the gradings of particular Lie algebras. References
\cite{sl4,sp4,o4} are the departure points of the present work.
Gradings of the exceptional Lie algebra $g_2$ are listed in
\cite{G2}, gradings of $f_4$ can be found in \cite{F4}.

\section{Preliminaries}\label{prel}

For a Lie algebra $(L,[\ ])$ we define a {\bf grading} $\Gamma$ as
a decomposition of the vector space $L$ into a direct sum of
subspaces $\{0\}\neq L_k\subseteq L$, $k\in\K$ fulfilling the
property $[L_k,L_l]\subseteq L_m$ for some $m\in\K$. It means, in
other words, that for each $k,l\in\K$ there exists $m\in\K$ such
that $[L_k,L_l]=\{[X_k,X_l]\,\vert\, X_k\in L_k, X_l\in
L_l\}\subseteq L_m$.

We denote such decomposition by $\Gamma :L=\oplus_{k\in\K}L_k$.

We call a grading $\widetilde{\Gamma}$ of $L$ a {\bf refinement}
of a grading $\Gamma :L=\oplus_{k\in\K}L_k$ if $\widetilde{\Gamma}
:L=\oplus_{k\in\K,i\in\I_k}L_{ki}$ and
$L_k=\oplus_{i\in\I_k}L_{ki}$ for each $k\in\K$.

By means of refinements, we can gradually split the gradings of
$L$ until it is impossible to make any further non-trivial
refinement. Such a grading whose each refinement is equal to
itself is called a fine grading. Overall, the set of all gradings
of the Lie algebra $L$ forms a structure of a 'tree', with $L$
itself on the top and the fine gradings of $L$ on the bottom.

Having a set of diagonalizable automorphisms that mutually
commute, it is possible to split the Lie algebra into subspaces
that are mutual eigensubspaces of all the automorphisms from this
set, thus obtaining a grading. Obviously, the bigger set of
automorphisms we use, the finer grading we receive by this
decomposition. Therefore, for studying the gradings of Lie
algebras, an important role is played by so-called MAD-groups. The
term {\bf MAD-group} of a Lie algebra $L$ is a short for maximal
Abelian subgroup of diagonalizable automorphisms (contained in the
group $\A ut\,L$ of all automorphisms on the Lie algebra $L$).

Indices of a grading obtained in this way can be embedded into a
group, thus enabling us to introduce the term of a group grading. We
say that the grading $\Gamma : L = \oplus_{k \in \K} L_k$ is a {\bf
group grading} if the index set $\K$ can be chosen as a subset of a
group in such a way that $\{0\}\neq[L_j,L_k]$ implies
$[L_j,L_k]\subseteq L_{j\star k}$. Note that for two different
grading subspaces $L_j$ and $L_k$ the corresponding group elements
$j$, $k$ must be different. If a refinement of a group gradings is
also a group grading, then we call it {\bf group refinement}. There
is a relationship between the group labelling the group refinement
and the group labelling the original group grading (see \cite{G2}).
More precisely,  the index set of a group grading can be embedded
into a universal Abelian group,  say $G$, such that the index set of
any coarsening is embedable into an image of $G$ by a group
epimorphism.

There is now a big confusion in literature between the terms grading
and group grading, caused by an incorrect statement from
\cite{OLG1}, which says that the index set of each grading of a Lie
algebra is embedable  into a semigroup. As already mentioned, there
exist counterexamples to this statement, some of them provided
firstly by \cite{Eld}. On the basis of that incorrect statement, the
following theorem was proved in \cite{OLG1}:

{\it For a simple Lie algebra $L$ over an algebraically closed
field, a grading $\Gamma$ of $L$ is fine if and only if there
exists a MAD-group $\G\subset\A ut\,L$ such that $\Gamma$ is a
decomposition of $L$ into simultaneous eigensubspaces of all
automorphisms from $\G$.}

It has not been shown yet whether this theorem is true, or not.

We have been often referring to this theorem in our previous works
(\cite{sl4,sp4,o4}), and concluding that for the Lie algebras
studied in there we have found all the fine gradings. In fact, we
may use only the following weaker statement about relation between
 MAD-groups and fine gradings.

\begin{thm}\label{new_1-1}
A group grading $\Gamma$ of a complex Lie algebra $L$ is fine, if
and only if there exists a MAD-group $\G\subset\A ut\,L$ such that
$\Gamma$ is obtained as a decomposition of $L$ into simultaneous
eigensubspaces of all automorphisms from $\G$.
\end{thm}

It means that, in the articles (\cite{sl4,sp4,o4}) cited above, we
have described all fine group gradings, but we cannot claim that
they are at the same time all the fine gradings. On the example of
the non-simple algebra $o(4,\C)$ we have found a fine grading which
is not a group grading (see the case of grading by the MAD-group
$\G_5$ in Table \ref{complex-o}). If we would try to find the
indices of the subspaces in the fine grading in order to obtain a
group grading, then two of the grading subspaces would have to be
labelled by the same index, thus giving rise to a coarser group
grading. Note that this happens for any (non-simple) Lie algebra $L$
in the form $L=L_1 \otimes L_2$, where the grading arises as a
composition of the two root gradings of the individual algebras
$L_1$ and $L_2$. These findings lead us to the following assumption:

\begin{conj}\label{gradings-simple}
On a simple complex Lie algebra, the terms fine grading and fine
group grading coincide.
\end{conj}

To summarize, we have a one-to-one correspondence between the fine
group gradings and the MAD-groups for complex Lie algebras. And
since the MAD-groups were fully classified for all the 'standard'
complex Lie algebras (meaning the classes $A_n$, $B_n$, $C_n$,
$D_n$, with the exception of $D_4$), we can obtain all the fine
group gradings for these algebras. However, we cannot conclude
that these fine group gradings are at the same time all the fine
gradings of the particular algebras, although we expect  that it
is the case for simple algebras.

For a grading $\Gamma:L=\oplus_{k\in\K}L_k$ of $L$ and an
automorphism $g\in\A ut\,L$, the decomposition
$\widetilde{\Gamma}:L=\oplus_{k\in\K}(gL_k)$ is also a grading of
$L$; and we call such gradings $\Gamma$ and $\widetilde{\Gamma}$
equivalent. Further on, when speaking about two 'different'
gradings, we mean non-equivalent gradings. Consequently,  if
$\Gamma$ is a group grading, then the equivalent grading
$\widetilde{\Gamma}$ is again a group grading with the same index
set $\K$ lying in the same group.

For real Lie algebras, however, we do not possess any simple tool
like MAD-groups. Our aim in this work is to describe fine group
gradings of all the real forms of the Lie algebras $sl(4,\C)$,
$sp(4,\C)$, and $o(4,\C)$, and relationships between them. For
classical simple complex Lie algebras, these relationships are
very straightforward, given the fact that $sp(4,\C)$ and $o(4,\C)$
are subalgebras of $sl(4,\C)$. It was proved in \cite{OLG2} that a
fine group grading of $sp(n,\C)$ or $o(n,\C)$, $n\neq 8$, is
always formed as a selection of several grading subspaces from a
fine group grading of $sl(n,\C)$. We will use this method also for
finding fine group gradings of real forms of the Lie algebras in
question.

Firstly, let us recall previous results (\cite{sl4,sp4,o4})
containing fine group gradings of the complex Lie algebras
$sl(4,\C)$, $sp(4,\C)$, and $o(4,\C)$ themselves, because they are
the starting point for exploration of fine group gradings of their
real forms. Then we shall move on to the core of this work -
namely the fine group gradings of the real forms of the three Lie
algebras in question. Note that gradings of $sp(4,\C)$ are also
studied in \cite{Bahturin}.

\section{Fine Group Gradings of the Complex Lie Algebras $sl(4,\C)$,
$sp(4,\C)$, and $o(4,\C)$}\label{complex}

\subsection{Fine Group Gradings of the Complex Lie Algebra $sl(4,\C)$}
\label{complex-sl}

Let us start with the algebra $sl(4,\C)=\{X\in\C^{4\times
4}\,\vert\, \tr(X)=\sum_{j=1}^4 X_{jj}=0\}$. On $sl(4,\C)$, there
exist two types of automorphisms:
\begin{itemize}
    \item inner automorphism: $Ad_A$ for $A\in Gl(4,\C)$, where
    $Ad_A (X)=A^{-1}XA$;
    \item outer automorphism: $Out_C$ for $C\in Gl(4,\C)$, where
    $Out_C (X)=-(C^{-1}XC)^T$.
\end{itemize}

The full set of automorphisms on $sl(4,\C)$ is then $$ \A ut\,
sl(4,\C)=\{Ad_A\,\vert\, \mathrm{det}\,A\neq
0\}\cup\{Out_C\,\vert\, \mathrm{det}\,C\neq 0\}.$$

In our list of MAD-groups, we always characterize each MAD-group
by the set $G_{Ad}$ of matrices $A$ for the inner automorphisms,
and by the set $G_{Out}$ of matrices $C$ for the outer
automorphisms if the MAD-group contains outer automorphisms at
all.

In fact, we provide just one element $C$ of $G_{Out}$, because any
other outer automorphism $Out_{\widetilde{C}}$ from the MAD-group is
a composition of the one $Out_C$ and of some inner automorphism
$Ad_A$ with $A\in G_{Ad}$.

There are eight MAD-groups on $sl(4,\C)$ listed in Table
\ref{MAD-complex-sl}; two of them with inner automorphisms only,
and six of them containing also outer automorphisms. The symbol
$\otimes$ used  in the Table stands for the tensor product of
matrices.\footnote{If $A\in\mathbb{C}^{n\times n}$ and  $B \in
\mathbb{C}^{m\times m}$, then the tensor product $A\otimes B \in
\mathbb{C}^{nm\times nm}$  is defined by $(A\otimes B)_{IJ}=
A_{i_1i_2}B_{j_1j_2}$, where $i_1,i_2 \in \{0,1,\ldots, n-1\}$,
$j_1,j_2 \in \{0,1,\ldots , m-1\}$, $I,J\in \{0,1,\ldots ,mn-1\}$
and $I=i_1m+j_1$, $J=i_2m+j_2$.}

\begin{table}[ht]
\begin{center}
\begin{tabular}{|l||l|l|}
\hline

&$G_{Ad}$&$G_{Out}$\\ \hline \hline

$\mathcal{G}_{1}$&$\{A=\diag\,(d_1, d_2, d_3, 1),\,d_j\in\C,
d_j\neq0\}$ &$\emptyset$\\ \hline

$\mathcal{G}_{2}$&$\{A=P^j Q^k,\,j,k=0,1,2,3,$&$\emptyset$\\
&$P=\left(\begin{smallmatrix}
                  0&1&0&0\\
                  0&0&1&0\\
                  0&0&0&1\\
                  1&0&0&0
\end{smallmatrix}\right),
Q=\diag\,(1,i, -1, -i)\}$&\\ \hline

$\mathcal{G}_{3}$&$\{A=\diag\,(\varepsilon_{1},
         \varepsilon_{2},
         \varepsilon_{3},
         1),\,\varepsilon_{i}=\pm 1\}$&$C=I_4$\\
\hline

$\mathcal{G}_{4}$&$\{A=\diag\,(1,
         \varepsilon,
         \alpha,
         \alpha^{-1}),\,\varepsilon=\pm 1,\alpha\in\C, \alpha\neq 0\}$
&$C=\left(\begin{smallmatrix}
            1&0&0&0\\
            0&1&0&0\\
            0&0&0&1\\
            0&0&1&0
            \end{smallmatrix}\right)$\\
\hline

$\mathcal{G}_{5}$&$\{A=\diag\,(\alpha,
         \alpha^{-1},\beta,\beta^{-1}),\,\alpha,\beta\in\C,
     \alpha, \beta\neq 0\}$
&$C=\left(\begin{smallmatrix}
            0&1&0&0\\
            1&0&0&0\\
            0&0&0&1\\
            0&0&1&0
            \end{smallmatrix}\right)$\\
\hline

$\mathcal{G}_{6}$&$\{A=\sigma_j\otimes \diag\,(\alpha,
         \alpha^{-1}),\,\alpha\in\C, \alpha\neq 0, j=0,1,2,3\}$
&$C=\left(\begin{smallmatrix}
            0&1&0&0\\
            1&0&0&0\\
            0&0&0&1\\
            0&0&1&0
            \end{smallmatrix}\right)$\\
\hline

$\mathcal{G}_{7}$&$\{A=\sigma_j\otimes\sigma_k,\,j,k=0,1,2,3\}$
&$C=I_4$\\ \hline

$\mathcal{G}_{8}$&$\{A\in(G_0\otimes I_2)\cup
                         (RG_0\otimes\sigma_3)\cup
             (G_0\otimes\sigma_1)\cup
                         (RG_0\otimes\sigma_2),$
        &$C=\left(\begin{smallmatrix}
            1&0&0&0\\
            0&1&0&0\\
            0&0&0&1\\
            0&0&1&0
            \end{smallmatrix}\right)$\\
                 &$G_0=\{\left(\begin{smallmatrix}1&0\\0&1
                 \end{smallmatrix}\right),
                 \left(\begin{smallmatrix}1&0\\0&-1
                 \end{smallmatrix}\right)\},
         R=\left(\begin{smallmatrix}1&0\\0&i
                 \end{smallmatrix}\right)
                 \}$&\\
\hline
\end{tabular}
\medskip
\caption{The eight non-conjugate MAD-groups on $sl(4,\C)$. The
symbols $\sigma_j$ stand for the $2\times 2$ real Pauli matrices
.}\label{MAD-complex-sl}
\end{center}
\end{table}

Non-conjugate MAD-groups generate non-equivalent fine group
gradings of the Lie algebra. For $sl(4,\C)$, the Theorem
\ref{new_1-1} implies that the eight fine group gradings generated
by the MAD-groups $\G_1, \ldots, \G_8$ form the full set of
non-equivalent fine group gradings on $sl(4,\C)$.

In the following tables, we provide an explicit summary of all
these eight fine group gradings, together with the universal
Abelian groups for their index sets. We will use the notation of
\begin{itemize}
    \item $\underline{L_k\subset sl(4,\C)}$: grading subspace $L_k$
    of a grading $\Gamma$ of $sl(4,\C)$ described as a
    complex span of the relevant basis vectors $X_i$, namely
    $L_k=\C\cdot X_{i_1}+\ldots+\C\cdot X_{i_k}= \mathrm{span}^{\C}(X_{i_1},\ldots, X_{i_k})$;
    \item $\underline{E_{jk}\in\C^{4\times 4}}$: matrices containing
    just one non-zero element and fifteen zeros, namely $(E_{jk})_{lm}=\delta_{jl}\delta_{km}$,
    where $\delta_{hi}$ is the Kronecker symbol.
\end{itemize}

\begin{table}[ht]
\begin{center}
\begin{tabular}{|c@{\,:\,}l|c@{\,:\,}l|c@{\,:\,}l|c@{\,:\,}l|}\hline
$L_1$&$X_1=E_{23}$&$L_2$&$X_2=E_{12}$&$L_3$&$X_3=E_{13}$&$L_4$&$X_4=E_{34}$\\
\hline
$L_5$&$X_5=E_{24}$&$L_6$&$X_6=E_{41}$&$L_7$&$X_7=E_{14}$&$L_8$&$X_8=E_{42}$\\
\hline
$L_9$&$X_9=E_{43}$&$L_{10}$&$X_{10}=E_{31}$&$L_{11}$&$X_{11}=E_{21}$&$L_{12}$&$X_{12}=E_{32}$\\
\hline $L_{13}$&\multicolumn{7}{l|}{$X_{13}=E_{11}-E_{22}\quad
X_{14}=E_{22}-E_{33}\quad X_{15}=E_{33}-E_{44}$}\\ \hline

\end{tabular}
\medskip
\caption{Fine group grading $\Gamma_1
:sl(4,\C)=\oplus_{k=1}^{13}L_k$ generated by MAD-group $\G_1$,
known as the 'root grading'. There are twelve one-dimensional
subspaces $L_k=\C\cdot X_k$, $k=1,\ldots,12$, and one
three-dimensional grading subspace $L_{13}=\C\cdot X_{13}+\C\cdot
X_{14}+\C\cdot X_{15}$. The universal Abelian group of this fine
group grading is $\Z^3$.}\label{G1}
\end{center}
\end{table}

\begin{table}[ht]
\begin{center}
\begin{tabular}{|c@{\,:\,}l|c@{\,:\,}l|c@{\,:\,}l|c@{\,:\,}l|c@{\,:\,}l|}\hline
$L_1$&$X_1=P^0Q^3$&$L_2$&$X_2=P^0Q^2$&$L_3$&$X_3=P^0Q^1$&
    $L_4$&$X_4=P^1Q^0$&$L_5$&$X_5=P^1Q^3$\\ \hline
$L_6$&$X_6=P^1Q^2$&$L_7$&$X_7=P^1Q^1$&$L_8$&$X_8=P^2Q^0$&
    $L_9$&$X_9=P^2Q^3$&$L_{10}$&$X_{10}=P^2Q^2$\\ \hline
$L_{11}$&$X_{11}=P^2Q^1$&$L_{12}$&$X_{12}=P^3Q^0$&$L_{13}$&$X_{13}=P^3Q^3$&
    $L_{14}$&$X_{14}=P^3Q^2$&$L_{15}$&$X_{15}=P^3Q^1$\\ \hline
\end{tabular}
\medskip
\caption{Fine group grading $\Gamma_2
:sl(4,\C)=\oplus_{k=1}^{15}L_k$ generated by MAD-group $\G_2$,
known as the 'Pauli grading'. All the grading subspaces are
one-dimensional: $L_k=\C\cdot X_k$, $k=1,\ldots,15$; the basis
matrices $X_k$ are expressed by means of the $4\times 4$ Pauli
matrices $P$, $Q$ defined in Table \ref{MAD-complex-sl}. The
universal Abelian group of this fine group grading is
$\Z_4^2$.}\label{G2}
\end{center}
\end{table}

\begin{table}[ht]
\begin{center}
\begin{tabular}{|c@{\,:\,}l|c@{\,:\,}l|c@{\,:\,}l|}\hline
$L_1$&$X_1=E_{14}+E_{41}$&$L_2$&$X_2=E_{14}-E_{41}$&$L_3$&$X_3=E_{24}+E_{42}$\\
\hline
$L_4$&$X_4=E_{24}-E_{42}$&$L_5$&$X_5=E_{34}+E_{43}$&$L_6$&$X_6=E_{34}-E_{43}$\\
\hline
$L_7$&$X_7=E_{13}+E_{31}$&$L_8$&$X_8=E_{13}-E_{31}$&$L_9$&$X_9=E_{23}+E_{32}$\\
\hline
$L_{10}$&$X_{10}=E_{23}-E_{32}$&$L_{11}$&$X_{11}=E_{12}+E_{21}$&$L_{12}$&$X_{12}=E_{12}-E_{21}$\\
\hline $L_{13}$&\multicolumn{5}{l|}{$X_{13}=E_{11}-E_{22} \quad
X_{14}=E_{22}-E_{33} \quad X_{15}=E_{33}-E_{44}$}\\ \hline
\end{tabular}
\medskip
\caption{Fine group grading $\Gamma_3
:sl(4,\C)=\oplus_{k=1}^{13}L_k$ generated by MAD-group $\G_3$.
There are twelve one-dimensional subspaces $L_k=\C\cdot X_k$,
$k=1,\ldots,12$, and one three-dimensional grading subspace
$L_{13}=\C\cdot X_{13}+\C\cdot X_{14}+\C\cdot X_{15}$. The
universal Abelian group of this fine group grading is
$\Z_2^4$.}\label{G3}
\end{center}
\end{table}

\begin{table}[ht]
\begin{center}
\begin{tabular}{|c@{\,:\,}l|c@{\,:\,}l|c@{\,:\,}l|}\hline
$L_1$&$X_1=E_{13}-E_{41}$&$L_2$&$X_2=E_{13}+E_{41}$&$L_3$&$X_3=E_{14}-E_{31}$\\
\hline
$L_4$&$X_4=E_{14}+E_{31}$&$L_5$&$X_5=E_{24}-E_{32}$&$L_6$&$X_6=E_{24}+E_{32}$\\
\hline
$L_7$&$X_7=E_{23}-E_{42}$&$L_8$&$X_8=E_{23}+E_{42}$&$L_9$&$X_9=E_{12}-E_{21}$\\
\hline
$L_{10}$&$X_{10}=E_{12}+E_{21}$&$L_{11}$&$X_{11}=E_{34}$&$L_{12}$&$X_{12}=E_{43}$\\
\hline
$L_{13}$&$X_{13}=E_{33}-E_{44}$&$L_{14}$&\multicolumn{3}{l|}{$X_{14}=E_{11}-E_{22}\quad
X_{15}=E_{11}+E_{22}-E_{33}-E_{44}$}\\ \hline
\end{tabular}
\medskip
\caption{Fine group grading $\Gamma_4
:sl(4,\C)=\oplus_{k=1}^{14}L_k$ generated by MAD-group $\G_4$.
Thirteen grading subspaces are one-dimensional: $L_k=\C\cdot X_k$,
$k=1,\ldots,13$, one is two-dimensional: $L_{14}=\C\cdot
X_{14}+\C\cdot X_{15}$. The universal Abelian group of this fine
group grading is $\Z\times\Z_2^2$.}\label{G4}
\end{center}
\end{table}

\begin{table}[ht]
\begin{center}
\begin{tabular}{|c@{\,:\,}l|c@{\,:\,}l|c@{\,:\,}l|}\hline
$L_1$&$X_1=E_{23}-E_{41}$&$L_6$&$X_6=E_{13}+E_{42}$&$L_{11}$&$X_{11}=E_{34}$\\
\hline
$L_2$&$X_2=E_{23}+E_{41}$&$L_7$&$X_7=E_{14}-E_{32}$&$L_{12}$&$X_{12}=E_{43}$\\
\hline
$L_3$&$X_3=E_{24}-E_{31}$&$L_8$&$X_8=E_{14}+E_{32}$&$L_{13}$&$X_{13}=E_{11}+E_{22}-E_{33}-E_{44}$\\
\hline
$L_4$&$X_4=E_{24}+E_{31}$&$L_9$&$X_9=E_{12}$&$L_{14}$&$X_{14}=E_{11}-E_{22}+E_{33}-E_{44}$\\
\cline{1-4}
$L_5$&$X_5=E_{13}-E_{42}$&$L_{10}$&$X_{10}=E_{21}$&\multicolumn{2}{l|}{$\qquad\
X_{15}=E_{11}-E_{22}-E_{33}+E_{44}$}\\ \hline
\end{tabular}
\medskip
\caption{Fine group grading $\Gamma_5
:sl(4,\C)=\oplus_{k=1}^{14}L_k$ generated by MAD-group $\G_5$.
There are thirteen one-dimensional grading subspaces: $L_k=\C\cdot
X_k$, $k=1,\ldots,13$, and one two-dimensional grading subspace:
$L_{14}=\C\cdot X_{14}+\C\cdot X_{15}$. The universal Abelian
group of this fine group grading is $\Z^2\times\Z_2$.}\label{G5}
\end{center}
\end{table}

\begin{table}[ht]
\begin{center}
\begin{tabular}{|c@{\,:\,}l|c@{\,:\,}l|c@{\,:\,}l|}\hline
$L_1$&$X_1=E_{11}-E_{22}+E_{33}-E_{44}$&$L_6$&$X_6=E_{13}+E_{24}+E_{31}+E_{42}$&$L_{11}$&$X_{11}=E_{14}+E_{32}$\\
\hline
$L_2$&$X_2=E_{11}-E_{22}-E_{33}+E_{44}$&$L_7$&$X_7=E_{13}-E_{24}-E_{31}+E_{42}$&$L_{12}$&$X_{12}=E_{23}-E_{41}$\\
\hline
$L_3$&$X_3=E_{11}+E_{22}-E_{33}-E_{44}$&$L_8$&$X_8=E_{21}+E_{43}$&$L_{13}$&$X_{13}=E_{23}+E_{41}$\\
\hline
$L_4$&$X_4=E_{13}-E_{24}+E_{31}-E_{42}$&$L_9$&$X_9=E_{21}-E_{43}$&$L_{14}$&$X_{14}=E_{12}+E_{34}$\\
\hline
$L_5$&$X_5=E_{13}+E_{24}-E_{31}-E_{42}$&$L_{10}$&$X_{10}=E_{14}-E_{32}$&$L_{15}$&$X_{15}=E_{12}-E_{34}$\\
\hline
\end{tabular}
\medskip
\caption{Fine group grading $\Gamma_6
:sl(4,\C)=\oplus_{k=1}^{15}L_k$ generated by MAD-group $\G_6$. All
of the grading subspaces are one-dimensional: $L_k=\C\cdot X_k$,
$k=1,\ldots,15$. The universal Abelian group of this fine group
grading is $\Z\times\Z_2^3$.}\label{G6}
\end{center}
\end{table}

\begin{table}[ht]
\begin{center}
\begin{tabular}{|c@{\,:\,}l|c@{\,:\,}l|}\hline
$L_1$&$X_1=E_{14}+E_{23}+E_{32}+E_{41}$&$L_9$&$X_9=E_{12}+E_{21}+E_{34}+E_{43}$\\
\hline
$L_2$&$X_2=E_{14}-E_{23}-E_{32}+E_{41}$&$L_{10}$&$X_{10}=E_{12}-E_{21}-E_{34}+E_{43}$\\
\hline
$L_3$&$X_3=E_{14}-E_{23}+E_{32}-E_{41}$&$L_{11}$&$X_{11}=E_{12}-E_{21}+E_{34}-E_{43}$\\
\hline
$L_4$&$X_4=E_{14}+E_{23}-E_{32}-E_{41}$&$L_{12}$&$X_{12}=E_{12}+E_{21}-E_{34}-E_{43}$\\
\hline
$L_5$&$X_5=E_{13}+E_{24}+E_{31}+E_{42}$&$L_{13}$&$X_{13}=E_{11}-E_{22}-E_{33}+E_{44}$\\
\hline
$L_6$&$X_6=E_{13}-E_{24}-E_{31}+E_{42}$&$L_{14}$&$X_{14}=E_{11}-E_{22}+E_{33}-E_{44}$\\
\hline
$L_7$&$X_7=E_{13}-E_{24}+E_{31}-E_{42}$&$L_{15}$&$X_{15}=E_{11}+E_{22}-E_{33}-E_{44}$\\
\hline $L_8$&$X_8=E_{13}+E_{24}-E_{31}-E_{42}$\\ \cline{1-2}
\end{tabular}
\medskip
\caption{Fine group grading $\Gamma_7
:sl(4,\C)=\oplus_{k=1}^{15}L_k$ generated by MAD-group $\G_7$. All
of the grading subspaces are one-dimensional: $L_k=\C\cdot X_k$,
$k=1,\ldots,15$. The universal Abelian group of this fine group
grading is $\Z_2^5$.}\label{G7}
\end{center}
\end{table}

\begin{table}[ht]
\begin{center}
\begin{tabular}{|c@{\,:\,}l|c@{\,:\,}l|c@{\,:\,}l|}\hline
$L_1$&$X_1=E_{13}+E_{24}+E_{32}+E_{41}$&$L_6$&$X_6=E_{14}-E_{23}-E_{31}+E_{42}$&$L_{11}$&$X_{11}=E_{33}-E_{44}$\\
\hline
$L_2$&$X_2=E_{13}-E_{24}-E_{32}+E_{41}$&$L_7$&$X_7=E_{14}-E_{23}+E_{31}-E_{42}$&$L_{12}$&$X_{12}=E_{12}-E_{21}$\\
\hline
$L_3$&$X_3=E_{13}-E_{24}+E_{32}-E_{41}$&$L_8$&$X_8=E_{14}+E_{23}-E_{31}-E_{42}$&$L_{13}$&$X_{13}=E_{34}-E_{43}$\\
\hline
$L_4$&$X_4=E_{13}+E_{24}-E_{32}-E_{41}$&$L_9$&$X_9=E_{11}+E_{22}-E_{33}-E_{44}$&$L_{14}$&$X_{14}=E_{12}+E_{21}$\\
\cline{1-4}
$L_5$&$X_5=E_{14}+E_{23}+E_{31}+E_{42}$&$L_{10}$&$X_{10}=E_{11}-E_{22}$&\multicolumn{2}{l|}{$\qquad\
X_{15}=E_{34}+E_{43}$}\\ \hline
\end{tabular}
\medskip
\caption{Fine group grading $\Gamma_8
:sl(4,\C)=\oplus_{k=1}^{14}L_k$ generated by MAD-group $\G_8$.
There are thirteen one-dimensional grading subspaces: $L_k=\C\cdot
X_k$, $k=1,\ldots,13$, and one two-dimensional grading subspace:
$L_{14}=\C\cdot X_{14}+\C\cdot X_{15}$. The universal Abelian
group of this fine group grading is $\Z_4\times\Z_2^2$.}\label{G8}
\end{center}
\end{table}

\subsection{Fine Group Gradings of the Complex Lie Algebras $sp(4,\C)$ and $o(4,\C)$}
\label{complex-sp+o}
Let us now proceed with the Lie algebras $sp(4,\C)$ and $o(4,\C)$.
Both these algebras are subalgebras of $sl(4,\C)$, and they have
various representations determined by non-singular matrices
$K\in\C^{4\times 4}$, with $K=-K^T$ for $sp_K(4,\C)$, and $K=K^T$
for $o_K(4,\C)$: $$sp_K(4,\C)=\{X\in\C^{4\times
4}\,\vert\,XK=-KX^T\},\ \mathrm{det}\,K\neq 0,\ K=-K^T,$$
$$o_K(4,\C)=\{X\in\C^{4\times 4}\,\vert\,XK=-KX^T\},\
\mathrm{det}\,K\neq 0,\ K=K^T.$$

We have two options how to find the fine group gradings of these
two Lie algebras:
\begin{itemize}
    \item \underline{'MAD-group' method}: Being Lie algebras
    themselves, $sp_K(4,\C)$ and $o_K(4,\C)$ have MAD-groups in
    their automorphism groups. We can use these MAD-groups for
    splitting the algebras into simultaneous eigensubspaces, thus
    obtaining fine group gradings.
    \item \underline{'Displayed' method}: As each fine group grading of
    $sp_K(4,\C)$ and $o_K(4,\C)$ can be extended to a fine group grading
    of $sl(4,\C)$, we have the possibility to find the fine group
    gradings of $sp_K(4,\C)$ and $o_K(4,\C)$ by means of the
    'Displayed' method. We take the fine group gradings of $sl(4,\C)$ and
    select some of the grading subspaces, so that they make up the
    respective subalgebra, already being in the form of a grading.
\end{itemize}

The success of the 'Displayed' method is ensured by the following
statement (proved in \cite{OLG2}):

\begin{thm}\label{Out_K}
Let $\Gamma$ be a fine group grading of $sl(n,\C)$ generated by
MAD-group $\G$. This grading $\Gamma$ displays the subalgebra
$o_K(n,\C)$ for $n\neq 8$, or $sp_K(n,\C)$, for matrices verifying
$K=K^T$, or $K=-K^T$ respectively, if and only if $\G$ contains the
automorphism $Out_K$.
\end{thm}

Practically, the usage of Theorem \ref{Out_K} reduces into
choosing those grading subspaces from the grading $\Gamma$ which
are eigensubspaces of this automorphism $Out_K$ corresponding to
eigenvalue $+1$. Indeed, this property is equivalent to being an
element of $sp_K(n,\C)$, $o_K(n,\C)$ respectively: $$ X=Out_K
X=-(K^{-1}XK)^T \qquad \Leftrightarrow \qquad XK=-KX^T.$$

The two methods ('MAD-group' and 'Displayed') lead in principle to
the same result. The advantage of the 'Displayed' method is that
the grading subspaces of $sp_K(4,\C)$ coincide with chosen grading
subspaces of $sl(4,\C)$, the same goes for $o_K(4,\C)$. This
comfort is paid a price for, though, because the fine gradings are
then expressed in various representations of $o_K(4,\C)$. (For
$sp_K(4,\C)$, we luckily happen to obtain all the fine group
gradings in the same representation even by the 'Displayed'
method.)

We obtain all the three fine group gradings of $sp_K(4,\C)$ and
all the six fine group gradings of $o_K(4,\C)$. All these fine
group gradings found in this way on $sp_K(4,\C)$ and $o_K(4,\C)$
(up to one exception) are at the same time fine gradings, as is
implied by the dimensions of the grading subspaces. The
exceptional case is the fine group grading of $o_K(4,\C)$
displayed by the fine group grading $\Gamma_5$ of $sl(4,\C)$
generated by the MAD-group $\G_5$ (see Table \ref{complex-o}). It
is not a fine (non-group) grading of $o_K(4,\C)$, because it can
be further refined, but this refinement is not a group grading any
more.

In Tables \ref{complex-sp} and \ref{complex-o}, we list the fine
group gradings of $sp_K(4,\C)$ and $o_K(4,\C)$ in the form as
provided by the 'Displayed' method, because this notation will be
useful later on for description of fine group gradings of their
real forms. We only need to keep in mind that the fine group
gradings of $sl(4,\C)$ display the fine group gradings of
$o_K(4,\C)$ in several different representations - i.e. defined by
different matrices $K=K^T$.

\begin{table}[ht]
\begin{center}
\begin{tabular}{|c|c|l|}\hline
MAD-group&defining matrix&grading subspaces of $sl(4,\C)$ making
up the\\ on $sl(4,\C)$&$K=-K^T$&respective fine group grading of
$sp_K(4,\C)$\\ \hline \hline

$\G_5$&&$L_{14}\oplus L_2\oplus L_{11}\oplus L_3\oplus
L_{10}\oplus L_9\oplus L_5\oplus L_{12}\oplus L_{8}$\\
\cline{1-1}\cline{3-3}

$\G_6$&$K_0=J=\left(\begin{smallmatrix}
            0&-1&0&0\\
            1&0&0&0\\
            0&0&0&-1\\
            0&0&1&0\end{smallmatrix}\right)$
    &$L_1\oplus L_2\oplus L_8\oplus L_9\oplus L_{11}\oplus
    L_4\oplus L_5\oplus L_{13}\oplus L_{14}\oplus L_{15}$\\
            \cline{1-1}\cline{3-3}

$\G_7$&&$L_{13}\oplus L_{14}\oplus L_1\oplus L_3\oplus L_8\oplus
L_7\oplus L_{12}\oplus L_{11}\oplus L_{10}\oplus L_9$\\ \hline
\end{tabular}
\medskip
\caption{Fine group gradings of $sp_K(4,\C)$ as displayed by fine
group gradings of $sl(4,\C)$. In all the three cases, the algebra
$sp_K(4,\C)$ happens to be displayed in the same
representation.}\label{complex-sp}
\end{center}
\end{table}

\begin{table}[ht]
\begin{center}
\begin{tabular}{|c|c|l|}\hline
MAD-group&defining matrix&grading subspaces of $sl(4,\C)$ making
up\\ on $sl(4,\C)$&$K=K^T$&the respective fine group grading of
$o_K(4,\C)$\\ \hline \hline

$\G_3$&$K_1=I=\left(\begin{smallmatrix}
            1&0&0&0\\
            0&1&0&0\\
            0&0&1&0\\
            0&0&0&1\end{smallmatrix}\right)$
    &$L_2\oplus L_4\oplus L_6\oplus L_8\oplus L_{10}\oplus L_{12}$\\ \hline

$\G_4$&$K_2=\left(\begin{smallmatrix}
            1&0&0&0\\
            0&1&0&0\\
            0&0&0&1\\
            0&0&1&0\end{smallmatrix}\right)$
    &$L_{13}\oplus L_1\oplus L_5\oplus L_9\oplus L_7\oplus L_3$\\ \hline

&\multicolumn{2}{l|}{$\rightarrow$ fine group grading displayed by
the fine group grading $\Gamma_5$ of $sl(4,\C)$,}\\

&\multicolumn{2}{l|}{but not a fine grading: $L_{14}\oplus
L_1\oplus L_3\oplus L_5\oplus L_7$}\\

&\multicolumn{2}{l|}{$\rightarrow$ we split $L_{14}$ into
$L_{14}^1=\C\cdot X_{14}$, $L_{14}^2=\C\cdot X_{15}$}\\

$\G_5$&\multicolumn{2}{l|}{$\rightarrow$ we get a fine (non-group) grading:}\\
\cline{2-3}

&$K_3=\left(\begin{smallmatrix}
            0&1&0&0\\
            1&0&0&0\\
            0&0&0&1\\
            0&0&1&0\end{smallmatrix}\right)$&$L_{14}^1\oplus L_{14}^2\oplus L_1\oplus
L_3\oplus L_5\oplus L_7$\\ \hline

$\G_6$&$K_3=\left(\begin{smallmatrix}
            0&1&0&0\\
            1&0&0&0\\
            0&0&0&1\\
            0&0&1&0\end{smallmatrix}\right)$
    &$L_1\oplus L_2\oplus L_{10}\oplus L_4\oplus L_5\oplus L_{12}$\\ \hline

$\G_7$&$K_1=I=\left(\begin{smallmatrix}
            1&0&0&0\\
            0&1&0&0\\
            0&0&1&0\\
            0&0&0&1\end{smallmatrix}\right)$
    &$L_3\oplus L_4\oplus L_8\oplus L_6\oplus L_{11}\oplus L_{10}$\\ \hline

$\G_8$&$K_2=\left(\begin{smallmatrix}
            1&0&0&0\\
            0&1&0&0\\
            0&0&0&1\\
            0&0&1&0\end{smallmatrix}\right)$
    &$L_{11}\oplus L_4\oplus L_3\oplus L_{12}\oplus L_8\oplus L_6$\\ \hline
\end{tabular}
\medskip
\caption{Fine group gradings of $o_K(4,\C)$ as displayed by the
fine group gradings of $sl(4,\C)$, but in various representations
given by different symmetric matrices $K$. With the exception of
$\G_5$, we directly get fine gradings of $o_K(4,\C)$; for $\G_5$,
we need to split the two-dimensional grading subspace into two,
and only then the grading becomes fine.}\label{complex-o}
\end{center}
\end{table}

\section{Fine Group Gradings of the Real Forms of $sl(4,\C)$,
$sp(4,\C)$, and $o(4,\C)$}\label{real}
A real Lie subalgebra $\widetilde{L}$ of a complex Lie algebra $L$
is called a {\bf real form} of $L$ if each element $x\in L$ is
uniquely representable in the form $x=u+iv$, where $u, v\in
\widetilde{L}$. Not every complex Lie algebra has real forms. On
the other hand, a given complex Lie algebra may, in general, have
several non-isomorphic real forms.

Two real forms $\widetilde{L}_1$ and $\widetilde{L}_2$ of a Lie
algebra $L$ are said to be {\bf isomorphic} if
$g(\widetilde{L}_1)=\widetilde{L}_2$ for some automorphism
$g\in\A\,ut\ L$.

For the classical complex Lie algebras, we have the following
description of their real forms (see \cite{Jacobson}):

Every real form of the complex Lie algebra $L$ can be expressed as
$L_{\J}=\{X\in L\,\vert\,\J(X)=X\}$, where $\J$ is an involutive
antiautomorphism on $L$. Let us recall the meaning of the term
'involutive antiautomorphism':
\begin{itemize}
    \item involutive: $\J^2=Id$;
    \item antiautomorphism: $\J(\alpha X+Y)=\overline{\alpha}\J(X)+\J(Y)$
    for each $X,Y\in L$, $\alpha\in\C$.
\end{itemize}

This definition of the real forms, together with full
classification of the system of involutive antiautomorphisms,
provide a practical tool for working with real forms of the
complex Lie algebras we are interested in, namely $sl(4,\C)$,
$sp(4,\C)$, and $o(4,\C)$.

\subsection{Fine Group Gradings of the Real Forms of $sl(4,\C)$}
\label{real-sl}

Let us write down the classification of all the real forms of
$sl(4,\C)$. This complex Lie algebra has five non-isomorphic real
forms, and we list them in Table \ref{real_forms-sl} together with
the respective involutive antiautomorphisms that define them via
$L_{\J}=\{X\in L\,\vert\,\J(X)=X\}$.

We introduce a special notation for one involutive
antiautomorphism on $sl(4,\C)$ by $\J_0(X)=\overline{X}$. This
mapping $\J_0$ plays an important role, because any involutive
antiautomorphism on $sl(4,\C)$ can be made up as a composition of
$\J_0$ and some automorphism $h\in \A ut\, sl(4,\C)$.

\begin{table}[ht]
\begin{center}
\begin{tabular}{|c|l|l@{=}l|}\hline
&$\J=\J_0 Ad_F$&$sl(4,\R)$&$L_{\J}=L_{\J_0 Ad_F}=$\\ a)&\quad
$F\overline{F}=I$ (circular)&&$\{X\in sl(4,\C)\,\vert\,\J_0
Ad_F(X)=X\}=$\\ &&&$\{X\in sl(4,\C)\,\vert\, XF=F\overline{X}\}$\\
\hline \hline

&$\J=\J_0 Ad_F$&$su^*(4)$&$L_{\J}=L_{\J_0 Ad_F}=$\\ b)&\quad
$F\overline{F}=-I$ (anticircular)&&$\{X\in sl(4,\C)\,\vert\,\J_0
Ad_F(X)=X\}=$\\ &&&$\{X\in sl(4,\C)\,\vert\, XF=F\overline{X}\}$\\
\hline \hline

&$\J=\J_0 Out_E$&$su(4-r,r)$&$L_{\J}=L_{\J_0 Out_E}=$\\ c)&\quad
$E=E^+$ (hermitian)&&$\{X\in sl(4,\C)\,\vert\,\J_0
Out_E(X)=X\}=$\\ &\quad $\mathrm{sgn}(E)=(4-r,0,r)$&&$\{X\in
sl(4,\C)\,\vert\, XE=-EX^+\}$\\ \cline{2-4}
&\multicolumn{3}{c|}{$r=0,1,2$ $\rightarrow$ $su(4,0)$, $su(3,1)$,
$su(2,2)$}\\ \hline

\end{tabular}
\medskip
\caption{The five non-isomorphic real forms of $sl(4,\C)$ and
their defining involutive antiautomorphisms. There is just one
real form of type a), another one of type b), and three real forms
of type c). The notation $\mathrm{sgn}(E)$ stands for a triplet of
non-negative integers, where the first one is the number of
positive elements in the spectrum of $E$, second is the number of
zeros in the spectrum, and third is the number of negative
elements in the spectrum.}\label{real_forms-sl}
\end{center}
\end{table}

\medskip

Now we get to the point of searching for the fine group gradings
of these real forms. One straightforward idea how this could be
done is to use the same method as for the complex algebra - namely
take the MAD-groups of the real form and split the real form into
simultaneous eigensubspaces of all the automorphisms in the
MAD-group. Let us denote this process again as the 'MAD-group'
method. For this purpose, let us look at the overview of the
MAD-groups on the real forms of $sl(4,\C)$.

For a group $\H$ of automorphisms on a complex Lie algebra $L$, we
denote by $\H^{\R}=\{h\in\H\,\vert\,\sigma(h)\subset\R\}$ the
so-called {\bf real part} of the group $\H$ - i.e. all
automorphisms $h$ from $\H$ with real spectrum $\sigma(h)$.

Let $\G\subset\A\,ut\ L$ be a MAD-group on the complex Lie algebra
$L$, and let $\G^{\R}$ be its real part. We say that the set
$\G^{\R}$ is maximal if there exists no such MAD-group
$\widetilde{\G}$ on $L$ that $\G^{\R}$ is conjugate to some proper
subgroup of $\widetilde{\G}^{\R}$.

It is proved (in \cite{OLG3}) that each MAD-group $\F$ on a real
form of a classical simple Lie algebra $L$ is equal to the maximal
real part $\G^{\R}$ of some MAD-group $\G$ on $L$. (Obviously,
$\F$ is a subgroup of $\G^{\R}$ for some MAD-group $\G$ on $L$,
because all automorphisms $g\in\F$ can be uniquely extended from
$L_{\J}$ to $L$ by $g^{\C}(X+i Y)=g(X)+i g(Y)$ - such extensions
are diagonalizable, their spectrum is real, and they mutually
commute.)

So, we start with the eight MAD-groups on $sl(4,\C)$, and take
their real parts. Not all of the real parts are maximal:

\begin{itemize}
    \item $\G^{\R}_2$ is not maximal, because it is equal to a proper
    subgroup of $\G^{\R}_7$: $\G^{\R}_2\varsubsetneqq \G^{\R}_7$;
    \item $\G^{\R}_8$ is not maximal, because it is conjugate to a
    proper subgroup of $\G^{\R}_7$:\\
    $\widetilde{\G}^{\R}_8=f\G^{\R}_8f^{-1}\varsubsetneqq
    \G^{\R}_7$, with $f=Ad_S$, $S=
    \left(\begin{smallmatrix}
            1&0\\
            0&1\end{smallmatrix}\right)\oplus
    \left(\begin{smallmatrix}
            \frac{1+i}{2}&\frac{1-i}{2}\\
            \frac{1-i}{2}&\frac{1+i}{2}\end{smallmatrix}\right)$.
\end{itemize}

Therefore, the real parts of $\G^{\R}_2$ and $\G^{\R}_8$ are not
MAD-groups on any of the five real forms of $sl(4,\C)$.

The remaining complex MAD-groups $\G_1$, $\G_3$, $\G_4$, $\G_5$,
$\G_6$, $\G_7$ do have maximal real parts $\G^{\R}_1$,
$\G^{\R}_3$, $\G^{\R}_4$, $\G^{\R}_5$, $\G^{\R}_6$, $\G^{\R}_7$.
See the explicit overview of all the real parts $\G^{\R}_j$, $j=1,
\ldots, 8$ in Table \ref{MAD-real_parts-sl}.

\begin{table}[ht]
\begin{center}
\begin{tabular}{|l||l|l|}
\hline

&$G^{\R}_{Ad}$&$G^{\R}_{Out}$\\ \hline \hline

$\G^{\R}_{1}$&$\{A=\diag(d_1, d_2, d_3, 1),\,d_j\in\R, d_j\neq0\}$
&$\emptyset$\\ \hline

$\G^{\R}_{3}=\G_3$&$\{A=\diag(\varepsilon_{1},
         \varepsilon_{2},
         \varepsilon_{3},
         \varepsilon_{4}),\,\varepsilon_{i}=\pm 1\}$&$C=I_4$\\
\hline

$\G^{\R}_{4}$&$\{A=\diag(\varepsilon_1,
         \varepsilon_2,
         \alpha,
         \alpha^{-1}),\,\varepsilon_i=\pm 1,\alpha\in\R, \alpha\neq 0\}$
&$C=\left(\begin{smallmatrix}
            1&0&0&0\\
            0&1&0&0\\
            0&0&0&1\\
            0&0&1&0
            \end{smallmatrix}\right)$\\
\hline

$\G^{\R}_{5}$&$\{A=\diag(\alpha,
         \alpha^{-1},\beta,\beta^{-1}),$
&$C=\left(\begin{smallmatrix}
            0&1&0&0\\
            1&0&0&0\\
            0&0&0&1\\
            0&0&1&0
            \end{smallmatrix}\right)$\\
&$\,(\alpha,\beta\in\R)\vee(\alpha,\beta\in i\R),
     \alpha, \beta\neq 0\}$&\\ \hline

$\G^{\R}_{6}$&$\{A=\sigma_j\otimes \diag(\alpha,
         \alpha^{-1}), A=\sigma_j\otimes \diag(\alpha,
         -\alpha^{-1}),$
&$C=\left(\begin{smallmatrix}
            0&1&0&0\\
            1&0&0&0\\
            0&0&0&1\\
            0&0&1&0
            \end{smallmatrix}\right)$\\
&$\,\alpha\in\R, \alpha\neq 0, j=0,1,2,3\}$&\\ \hline

$\G^{\R}_{7}=\G_7$&$\{A=\sigma_j\otimes\sigma_k,\,j,k=0,1,2,3\}$
&$C=I_4$\\ \hline \hline

$\G^{\R}_{2}\varsubsetneqq\G^{\R}_7$&$\{A\in\{\sigma_0\otimes\sigma_0,
\sigma_1\otimes\sigma_0, \sigma_0\otimes\sigma_3,
\sigma_1\otimes\sigma_3\}\}$&$\emptyset$\\ \hline

$\widetilde{\G}^{\R}_{8}\varsubsetneqq\G^{\R}_7$&$\{A\in\{\sigma_0\otimes\sigma_0,
\sigma_3\otimes\sigma_0, \sigma_0\otimes\sigma_1,
\sigma_3\otimes\sigma_1\}\}$&$C=I_4$\\ \hline

\end{tabular}
\medskip
\caption{List of all non-conjugate real parts $\G^{\R}_j$ of
MAD-groups $\G_j$ on $sl(4,\C)$. The first six items are the
maximal ones ($\G^{\R}_1$, $\G^{\R}_3$, $\G^{\R}_4$, $\G^{\R}_5$,
$\G^{\R}_6$, $\G^{\R}_7$), while the last two are not maximal
($\G^{\R}_2$, and $\widetilde{\G}^{\R}_8$ - already in the
conjugate form, so as to see the link with $\G^{\R}_7$ easily).
The notation of $G_{Ad}$ and $G_{Out}$ is the same as introduced
in Table~\ref{MAD-complex-sl}.}\label{MAD-real_parts-sl}
\end{center}
\end{table}

From \cite{OLG3}, we have a result summarizing which of the real
forms $\G^{\R}_j$ are in fact MAD-groups on the individual real
forms of $sl(4,\C)$ - see the overview in Table \ref{MAD-real-sl}.

\begin{table}[ht]
\begin{center}
\begin{tabular}{|l|l|}
\hline real form &MAD-groups of the real form\\ \hline \hline
 $sl(4,\R)$&$\G_1^{\R}, \G_3^{\R}, \G_4^{\R}, \G_5^{\R},
\G_6^{\R}, \G_7^{\R}$\\ \hline
 $su^{*}(4)$&$\G_6^{\R}, \G_7^{\R}$\\ \hline
 $su(4,0)$&$\G_3^{\R}, \G_7^{\R}$\\ \hline
 $su(3,1)$&$\G_3^{\R}, \G_4^{\R}$\\ \hline
 $su(2,2)$&$\G_3^{\R}, \G_4^{\R}, \G_5^{\R},
\G_6^{\R}, \G_7^{\R}$\\ \hline
\end{tabular}
\medskip
\caption{MAD-groups on the five real forms of
$sl(4,\C)$}\label{MAD-real-sl}
\end{center}
\end{table}

Now that we dispose of the full list of MAD-groups on all the five
real forms, we can apply the 'MAD-group' method - i.e. decompose
the real forms into simultaneous eigensubspaces of all elements of
the respective MAD-groups, and by that obtain the fine group
gradings. We do not have here the one-to-one correspondence as
guaranteed for the complex Lie algebras by Theorem \ref{new_1-1},
however, we can at least make use of a weaker statement (proved in
\cite{OLG3}):

\begin{thm}\label{1-0}
\underline{'MAD-group' method:} Let $L_{\J}$ be a real form of a
classical complex Lie algebra $L$, $L\neq o(8,\C)$. Let $\F$ be a
MAD-group on $L_{\J}$. Then the grading of $L_{\J}$ generated by
$\F$ is a fine group grading.
\end{thm}

This theorem ensures that each MAD-group on the real form
generates a fine group grading, however, it does not ensure the
opposite implication - namely that this method provides all the
fine group gradings of the real forms. And indeed, we manage to
prove that there exist more fine group gradings of the real forms
of $sl(4,\C)$ than those generated by MAD-groups, simply by
finding counterexamples of additional fine group gradings that are
not generated by any MAD-group of the real form.

We do this by using the following method - let us call it the
'Fundamental' method. We work with the basis vectors of the
grading subspaces of the complex gradings and transform them, so
as to obtain a fine group grading of the real form; more
concretely:

\begin{thm}\label{fundamental}
\underline{'Fundamental' method:} Let
$\Gamma:L=\oplus_{k\in\K}L_k$ be a fine group grading of the
complex Lie algebra $L$. Let $\J$ be an involutive
antiautomorphism on $L$, let $L_{\J}$ be the real form of $L$
defined by $\J$, and let $Z_{k,l}$ be elements of $L_k$ fulfilling
the following properties:
\begin{itemize}
    \item $(Z_{k,1},\ldots,Z_{k,l_k})$ is a basis of the grading
    subspace $L_k$ for each $k\in\K$ - i.e. \newline
    $L_k = \mathrm{span}^{\C}(Z_{k,1},\ldots,Z_{k,l_k})$;
    \item $\J(Z_{k,l})=Z_{k,l}$ - i.e. $Z_{k,l}\in L_{\J}$ for all
    $Z_{k,l}\in L_k$.
\end{itemize}
Then the decomposition
$\Gamma^{\J}:L_{\J}=\oplus_{k\in\K}L^{\R}_k$, where
$L^{\R}_k=\mathrm{span}^{\R}(Z_{k,1},\ldots,Z_{k,l_k})$, is a fine
group grading of the real form $L_{\J}$.\\ We then say that the
grading $\Gamma$ of the complex Lie algebra $L$ {\bf determines}
the fine group grading $\Gamma^{\J}$ of the real form $L_{\J}$.
\end{thm}

Note that the same statement is valid also when considering fine
gradings instead of fine group gradings.

The assumptions imposed on the grading subspaces $L_k$ are
formulated as two conditions for the basis vectors of these
grading subspaces. This algorithmic approach can be modified into
just one requirement in the form $L_{\J} = \oplus_{k\in\K} L_k
\cap L_{\J}$. In the notation of Theorem \ref{fundamental}, the
subspace $L_k^{\R}$ equals to $L_k \cap L_{\J}$. This expression
clearly shows that the fine group grading of $L_{\J}$ obtained in
this way from the fine group grading of $L$ is unique.

Practically, the usage of this method consists in the following
process:
\begin{itemize}
    \item For a given fine group grading $\Gamma$ of $sl(4,\C)$, we
    search for an involutive antiautomorphism $h$ in the form $h=\J_0
    Out_E$ or $h=\J_0 Ad_F$, with matrices $E$, $F$ fulfilling the
    requirements listed in Table \ref{real_forms-sl}, such that
    there exists a basis $(Z_{k,1},\ldots,Z_{k,l_k})$ of $L_k$ with
    $\J(Z_{k,l})=Z_{k,l}$ - as required in Theorem
    \ref{fundamental}. If successful, this process results in
    finding the fine group gradings of as many of the real forms of $sl(4,\C)$ as
    possible.
    \item We repeat this process for all the fine group gradings
    $\Gamma_j$ of $sl(4,\C)$.
\end{itemize}

We need to keep in mind that various fine group gradings
$\Gamma_j$ of $sl(4,\C)$ determine fine group gradings of one real
form in different representations, depending on the defining
matrix $E$ or $F$ found during the process.

\medskip

As an illustration of the calculations carried out by the
'Fundamental' method, we give an example of how we found out which
fine group gradings of real forms of $sl(4,\C)$ are determined by
the fine grading $\Gamma_2$ of $sl(4,\C)$. Even though the real
part $\G_2^{\R}$ of the MAD-group $\G_2$, which generates
$\Gamma_2$, does not provide MAD-group on any of the real forms of
$sl(4,\C)$, we show that this grading $\Gamma_2$ does determine a
fine group grading on $su(3,1)$ and on $su(2,2)$, and does not
determine fine group gradings on the remaining real forms of
$sl(4,\C)$.

Firstly, let us inspect the cases $su(4-r,r)$. Since the grading
subspaces $L_k$ of the grading $\Gamma_2$ are one-dimensional, and
thus $L_k=\C\cdot X_k$, we need to find the convenient basis of
$L_k$ for the Theorem \ref{fundamental} in the form $Z_k=\alpha_k
X_k$, with $\alpha_k\in\C\setminus\{0\}$. We look for a hermitian
matrix $E$ ($E=E^+$) with $4-r$ positive and $r$ negative
eigenvalues such that, for all $X_k$ from the basis listed in
Table \ref{G2}, there exists $\alpha_k\in\C\setminus\{0\}$
fulfilling the equation
\begin{equation}\label{hledejE}
\J(\alpha_kX_k)=\J_0 Out_E(\alpha_kX_k)=\alpha_kX_k \qquad
\Leftrightarrow \qquad \alpha_kX_kE=-\overline{\alpha_k}EX^+_k.
\end{equation}
Let us start with $X_4=P=\left(\begin{smallmatrix}
         0&1&0&0\\0&0&1&0\\0&0&0&1\\1&0&0&0
      \end{smallmatrix}\right)$. Denoting
      $\eta=\frac{\overline{\alpha_4}}{\alpha_4}$, we obtain from
\eqref{hledejE} that the matrix $E$ has the form
$E=\left(\begin{smallmatrix}
         e_{11}&e_{12}&e_{13}&e_{14}\\
         -\eta e_{12}&-\eta e_{13}&-\eta e_{14}&-\eta e_{11}\\
         \eta^2 e_{13}&\eta^2 e_{14}&\eta^2 e_{11}&\eta^2 e_{12}\\
         -\eta^3 e_{14}&-\eta^3 e_{11}&-\eta^3 e_{12}&-\eta^3 e_{13}
      \end{smallmatrix}\right)$, and $\eta^4=1$.
We continue with $X_3=Q=\diag(1,i,-1,-i)$ and we find out that the
following four equations must hold so that the condition
\eqref{hledejE} could be fulfilled:
\begin{eqnarray}
(\alpha_3+\overline{\alpha_3})e_{11}&=&0 \nonumber\\
(\alpha_3-i\overline{\alpha_3})e_{12}&=&0 \label{eii!=0}\\
(\alpha_3-\overline{\alpha_3})e_{13}&=&0 \nonumber\\
(\alpha_3+i\overline{\alpha_3})e_{14}&=&0 \nonumber
\end{eqnarray}

This set of equations implies that at most one of the four
elements $e_{11}, e_{12}, e_{13}, e_{14}$ is non-zero. On the
other hand, at least one of them must be non-zero, otherwise the
matrix $E$ would be a zero matrix, and thus it would not have
$4-r$ positive and $r$ negative eigenvalues. So we have four
possibilities and we will go through them gradually:

\begin{itemize}

\item Let $e_{11}\neq 0$. It follows from \eqref{eii!=0} that
$e_{12}=e_{13}=e_{14}=0$. Then the matrix $E$ equals
$E=e_{11}\left(\begin{smallmatrix}
         1&0&0&0\\
         0&0&0&-\eta\\
         0&0&\eta^2&0\\
         0&-\eta^3&0&0
      \end{smallmatrix}\right)$. Spectrum of $E$ is
$\sigma(E)=\{e_{11},e_{11}, -e_{11}, \eta^2e_{11}\}$,
$e_{11}\in\R\setminus\{0\}$, since $E$ must be hermitian. So there
is at least one positive and one negative eigenvalue in
$\sigma(E)$. Thus $\J_0 Out_E$ with this type of $E$ does not
define the real form $su(4,0)$. But for $su(3,1)$ and $su(2,2)$ we
are successful:
    \begin{itemize}
    \item With $e_{11}=1$, $\alpha_4=i$, $\eta^2=1$ the spectrum of
    $E^{(3,1)}_3=\left(\begin{smallmatrix}
         1&0&0&0\\
         0&0&0&1\\
         0&0&1&0\\
         0&1&0&0
      \end{smallmatrix}\right)$ is equal to
      $\sigma(E^{(3,1)}_3)=\{1,1,1,-1\}$. Therefore, the fine group grading
      $\Gamma_2$ of $sl(4,\C)$ determines a fine group grading of
      $su(3,1)=L_{\J_0 Out_{E^{(3,1)}_3}}$.
    \item Putting $e_{11}=1$, $\alpha_4=1-i$, $\eta^2=-1$, the
    spectrum of $E^{(2,2)}_4=\left(\begin{smallmatrix}
         1&0&0&0\\
         0&0&0&-i\\
         0&0&-1&0\\
         0&i&0&0
      \end{smallmatrix}\right)$ is equal to
    $\sigma(E^{(2,2)}_4)=\{1,1,-1,-1\}$. Therefore, the fine group
    grading $\Gamma_2$ of $sl(4,\C)$ determines a fine group grading of
    $su(2,2)=L_{\J_0 Out_{E^{(2,2)}_4}}$.
    \end{itemize}
\item Let $e_{12}\neq 0$. Then $e_{11}=e_{13}=e_{14}=0$,
\begin{equation}\label{e12nenula}
E=e_{12}\left(\begin{smallmatrix}
         0&1&0&0\\
         -\eta&0&0&0\\
         0&0&0&\eta^2\\
         0&0&-\eta^3&0
      \end{smallmatrix}\right),
\end{equation}
and $\sigma(E)=\{\pm i\sqrt{\eta}e_{12}, \pm
i\sqrt{\eta}e_{12}\}$. Every hermitian matrix of this type has two
positive and two negative eigenvalues, but $su(2,2)$ is already
covered by the previous type. The important fact is that we cannot
obtain any hermitian matrix $E$ of the type \eqref{e12nenula} with
positive spectrum.

\item Let $e_{13}\neq 0$. Then $e_{11}=e_{12}=e_{14}=0$,
$E=e_{13}\left(\begin{smallmatrix}
         0&0&1&0\\
         0&-\eta&0&0\\
         \eta^2&0&0&0\\
         0&0&0&-\eta^3
      \end{smallmatrix}\right)$,
and $\sigma(E)=$ \newline $\{\pm \eta e_{13}, -\eta e_{13},
-\eta^3 e_{13}\}$. Again this kind of matrix $E$ does not admit
positive spectrum.

\item Let $e_{14}\neq 0$. Then $e_{11}=e_{12}=e_{13}=0$,
$E=e_{14}\left(\begin{smallmatrix}
         0&0&0&1\\
         0&0&-\eta&0\\
         0&\eta^2&0&0\\
         -\eta^3&0&0&0
      \end{smallmatrix}\right)$,
 $\sigma(E)=$ \newline $\{\pm i\eta\sqrt{\eta} e_{14}, \pm i\eta\sqrt{\eta}
 e_{14}\}$, which, again, does not admit positive spectrum for any
 matrix $E$.

\end{itemize}

One can see that none of the four possibilities allows a hermitian
matrix $E$ with positive spectrum. Therefore $\Gamma_2$ does not
determine fine group grading on any representation of the real
form
$su(4,0)$.\\

Now we move to the two real forms $sl(4,\R)$ and $su^*(4)$. The
real form $sl(4,\R)$ (resp. $su^*(4)$) has a fine group grading
determined by $\Gamma_2$ if there exists a circular (resp.
anticircular) matrix $F$ such that, for each $X_k$ from the basis
listed in Table \ref{G2}, there exists
$\alpha_k\in\C\setminus\{0\}$ fulfilling
\begin{equation}\label{hledejKzase}
\J(\alpha_kX_k)=\J_0 Ad_F(\alpha_kX_k)=\alpha_kX_k \qquad
\Leftrightarrow \qquad
\alpha_kX_kF=\overline{\alpha_k}F\overline{X_k}.
\end{equation}
For $X_4=P$ the condition \eqref{hledejKzase} implies that $F$ has
the form
\begin{equation}
F=\left(\begin{smallmatrix}
         f_{11}&f_{12}&f_{13}&f_{14}\\
         \nu f_{14}&\nu f_{11}&\nu f_{12}&\nu f_{13}\\
         \nu^2 f_{13}&\nu^2 f_{14}&\nu^2 f_{11}&\nu^2 f_{12}\\
         \nu^3 f_{12}&\nu^3 f_{13}&\nu^3 f_{14}&\nu^3 f_{11}
      \end{smallmatrix}\right),
\end{equation}
 where $\nu=\frac{\overline{\alpha_4}}{\alpha_4}$.
Again we proceed with inserting $X_3=Q$ into the equation
\eqref{hledejKzase}. Denoting
$\mu=\frac{\overline{\alpha_3}}{\alpha_3}$, we arrive at the
following set of equations:
\begin{equation}\label{mu}
\begin{array}{rclcrcl}
(1-\mu)f_{11}&=&0&\qquad\qquad&(1+\mu)f_{11}&=&0\\
(1-i\mu)f_{12}&=&0&\qquad\qquad&(1+i\mu)f_{12}&=&0\\
(1-\mu)f_{13}&=&0&\qquad\qquad&(1+\mu)f_{13}&=&0\\
(1-i\mu)f_{14}&=&0&\qquad\qquad&(1+i\mu)f_{14}&=&0.
\end{array}
\end{equation}
Clearly \eqref{mu} can only be fulfilled when
$f_{11}=f_{12}=f_{13}=f_{14}=0$. But it means that $F$ would be a
zero matrix. And a zero matrix $\Theta$ is not circular:
$F\overline{F}=\Theta\overline{\Theta}=\Theta\neq I$. Thus from
$\Gamma_2$ one does not obtain any fine group grading of
$sl(4,\R)$. The situation with $su^*(4)$ is analogous - by the
same process we come to the conclusion that $F=\Theta$, and a zero
matrix cannot be anticircular
($F\overline{F}=\Theta\overline{\Theta}=\Theta\neq -I$). Neither
$su^*(4)$ has any fine group grading determined by $\Gamma_2$.

\medskip

Apart from Theorem \ref{fundamental}, we can use one more powerful
tool, which simplifies the process of the 'Fundamental' method for
those fine gradings $\Gamma$ of $sl(4,\C)$ whose grading subspaces
have basis matrices with real elements only, and that is the case
for $\Gamma_j$, $j=1,3,4,5,6,7,8$.

\begin{thm}\label{real_basis}
\underline{'Real Basis' method:} Let $\G$ be a MAD-group on the
complex Lie algebra $sl(n,\C)$ and let $\Gamma
:sl(n,\C)=\oplus_{k\in\K}L_k$ be the fine group grading of
$sl(n,\C)$ generated by $\G$ such that all the subspaces
$L_k=\mathrm{span}^{\C}(X_{k,1},\ldots,X_{k,l_k})$ have real basis
vectors $X_{k,l}$. Let $h$ be an automorphism in $\A
ut\,sl(n,\C)$, such that $\J=\J_0h$ is an involutive
antiautomorphism on $sl(n,\C)$. Then the fine group grading
$\Gamma$ of $sl(n,\C)$ determines a fine group grading of the real
form $L_{\J}=L_{\J_0 h}$ if and only if the automorphism $h$ is an
element of the MAD-group $\G$.
\end{thm}

\begin{proof}
Firstly, let us consider an automorphism $h\in\G$. The
automorphism $h$ is either of the type $h=Ad_F$ with $F$ circular
or anticircular ($F\overline{F}=\pm I$), or $h=Out_E$ with $E$
hermitian ($E=E^+$). Since $\G$ is the MAD-group generating the
group grading $\Gamma$, any $X_{k,l}\in L_k$ is an eigenvector of
$h$. We show that, for both the types of $h$ and for any of the
basis vectors $X_{k,l}$, the eigenvalue $\lambda_{k}$ of the
eigenvector $X_{k,l}$ has its absolute value equal to $1$:
\begin{itemize}
    \item inner automorphism $h=Ad_F$:\\
    $\lambda X=h(X)=F^{-1}XF=\overline{F}X\overline{F}^{-1}=\overline{FXF^{-1}}=
    \overline{F(\frac{1}{\lambda}F^{-1}XF)F^{-1}}=
    \frac{1}{\overline{\lambda}}X$
    \item outer automorphism $h=Out_E$:\\
    $\lambda X=h(X)=-(E^{-1}XE)^T=-E^TX^TE^{-T}=
    -\overline{E^+X^T(E^+)^{-1}}=$\\
    \qquad\qquad $=-\overline{EX^TE^{-1}}=
    -\overline{E(-\frac{1}{\lambda}E^{-1}XE)E^{-1}}=
    \frac{1}{\overline{\lambda}}X$
\end{itemize}
The basis vectors $X_{k,l}$ are non-zero, therefore
$\lambda_{k}X_{k,l}=\frac{1}{\overline{\lambda_{k}}}X_{k,l}$
implies that $|\lambda_{k}|=1$. Thus we can express $\lambda_{k}$
as $\lambda_{k}=\cos\varphi_{k}+i\sin\varphi_{k}$. We define a new
basis vector $Z_{k,l}=\alpha_{k} X_{k,l}$, where
$\alpha_{k}=\cos(-\frac{\varphi_{k}}{2})+i\sin
(-\frac{\varphi_{k}}{2})$. (Such choice of $\alpha_{k}$ ensures
the property
$\lambda_{k}=\frac{\overline{\alpha_{k}}}{\alpha_{k}}$.) This
basis vector $Z_{k,l}$ is an element of $L_{\J_0 h}$: $\J_0
h(Z_{k,l})=\J_0 h(\alpha_{k}X_{k,l})= \J_0(\alpha_{k}h(X_{k,l})) =
\J_0(\alpha_{k}\lambda_{k}X_{k,l}) =
\J_0(\alpha_{k}\frac{\overline{\alpha_{k}}}{\alpha_{k}}X_{k,l}) =
\J_0(\overline{\alpha_{k}}X_{k,l})=\alpha_{k}\J_0(X_{k,l}) =
\alpha_{k}\overline{X_{k,l}} = \alpha_{k}X_{k,l}=Z_{k,l}.$ By
Theorem \ref{fundamental} ('Fundamental' method) it follows that
the sum $L_{\J_0 h}=\oplus_{k\in\K}L^{\R}_k$ with subspaces
$L^{\R}_k=\mathrm{span}^{\R}(Z_{k,1},\ldots,Z_{k,l_k})$ is a fine
group grading of $L_{\J_0 h}$.

Secondly, we assume that $\Gamma$ determines a fine group grading of
the real form $L_{\J} = L_{\J_0 h}$ with an automorphism $h\in\A
ut\, sl(n,\C)$. We start by showing that $h$ commutes with any
element of the MAD-group $\G$. Notice that $\J_0(L_k)=L_k$ by
hypothesis; and let $g\in\G$. It follows from the definition of the
grading $\Gamma$ that for any $k\in\K$ there exists $\alpha_k\in\C$
such that $g\mid_{L_k} =\alpha_k Id$. In order to prove that
$gh=gh$, it is sufficient to show that $hg\mid_{L_k}=gh\mid_{L_k}$
for any $k\in\K$. Clearly $L_{\J} = \oplus (L_k \cap L_{\J})$, and
therefore $\C\cdot(L_k \cap L_{\J}) = L_k$. Thus, for any $x\in L_k$
there are $y,z\in L_k \cap L_{\J}$ such that $x=y+iz$. In
particular, $\J(y)=y$ and $h(y)=\overline{y}$, and analogously for
$z$. It follows that $h(x) = h(y) + ih(z) = \overline{y}
+i\overline{z}$. Knowing that $\overline{y}, \overline{z}
\in\J_0(L_k) = L_k$, we apply $g$ and obtain $gh(x) =
\alpha_k(\overline{y-iz})$. Finally, the action of $g$ on $x$ by
$g(x)= \alpha_k x$ ensures that $hg(x) = h (\alpha_k x) = \alpha_k
h(x) = \alpha_k(\overline{y-iz})$, which means that $g$ and $h$
commute. Then, when checking in \cite{OLG2} the form of the
MAD-groups on $sl(n,\C)$, one derives that any automorphism which
commutes with the whole MAD-group must necessarily be
diagonalizable. And that already implies, due to the maximality of
$\G$, that $h\in\G$.
\end{proof}

\medskip

Let us now summarize the results of our calculations on the real
forms of $sl(4,\C)$; partially obtained with the use of Theorem
\ref{fundamental} - 'Fundamental' method (for $\Gamma_2$), partially
with the additional help of Theorem \ref{real_basis} - 'Real Basis'
method (for $\Gamma_8$). In Table
\ref{real_forms-sl-representations}, we list the various
representations of the real forms that will be used in the result
summary.

\begin{table}[ht]
\begin{center}
\begin{tabular}{|l|l|}
\hline Defining matrix $F^{(4,\R)}$&Representation of
$sl(4,\R)=L_{\J_0 Ad_{F^{(4,\R)}}}$\\ \hline
$F^{(4,\R)}_1=I_4$&$X=\left(\begin{smallmatrix}
            a&b&c&d\\
            e&f&g&h\\
            j&k&l&m\\
            n&o&p&-(a+f+l)
            \end{smallmatrix}\right)$\\ \hline \hline

Defining matrix $F^{^*(4)}$&Representation of $su^*(4)=L_{\J_0
Ad_{F^{^*(4)}}}$\\ \hline $F^{^*(4)}_1=\left(\begin{smallmatrix}
                    0&0&1&0\\
                    0&0&0&1\\
                    -1&0&0&0\\
                    0&-1&0&0
\end{smallmatrix}\right)$&
$X=\left(\begin{smallmatrix}
            a+il&b+im&c+ij&d+ik\\
            e+ip&-a+if&g+in&h+io\\
            -c+ij&-d+ik&a-il&b-im\\
            -g+in&-h+io&l-ip&-a-if
            \end{smallmatrix}\right)$\\ \hline
$F^{^*(4)}_2=\left(\begin{smallmatrix}
                    0&1&0&0\\
                    -1&0&0&0\\
                    0&0&0&i\\
                    0&0&-i&0
\end{smallmatrix}\right)$&
$X=\left(\begin{smallmatrix}
            a+if&b+ie&c+ih&d+ig\\
            -b+ie&a-if&id+g&-ic-h\\
            j+io&k+in&-a+il&m+ip\\
            -ik-n&ij+o&-m+ip&-a-il
            \end{smallmatrix}\right)$\\ \hline \hline

Defining matrix $E^{(4,0)}$&Representation of $su(4,0)=L_{\J_0
Out_{E^{(4,0)}}}$\\ \hline
$E^{(4,0)}_1=I_4$&$X=\left(\begin{smallmatrix}
            ia&b+ie&c+ij&d+in\\
            -b+ie&if&g+ik&h+io\\
            -c+ij&-g+ik&il&m+ip\\
            -d+in&-h+io&-m+ip&-i(a+f+l)
            \end{smallmatrix}\right)$\\ \hline \hline

Defining matrix $E^{(3,1)}$&Representation of $su(3,1)=L_{\J_0
Out_{E^{(3,1)}}}$\\ \hline $E^{(3,1)}_1=\left(\begin{smallmatrix}
                    1&0&0&0\\
                    0&1&0&0\\
                    0&0&1&0\\
                    0&0&0&-1
\end{smallmatrix}\right)$&
$X=\left(\begin{smallmatrix}
            ia&b+ie&c+ij&d+in\\
            -b+ie&if&g+ik&h+io\\
            -c+ij&-g+ik&il&m+ip\\
            d-in&h-io&m-ip&-i(a+f+l)
            \end{smallmatrix}\right)$\\ \hline
$E^{(3,1)}_2=\left(\begin{smallmatrix}
                    1&0&0&0\\
                    0&1&0&0\\
                    0&0&0&1\\
                    0&0&1&0
\end{smallmatrix}\right)$&
$X=\left(\begin{smallmatrix}
            ia&b+ie&c+in&d+ij\\
            -b+ie&if&g+io&h+ik\\
            -d+ij&-h+ik&l-\frac{i}{2}(a+f)&im\\
            -c+in&-g+io&ip&-l-\frac{i}{2}(a+f)
            \end{smallmatrix}\right)$\\ \hline
$E^{(3,1)}_3=\left(\begin{smallmatrix}
                    1&0&0&0\\
                    0&0&0&1\\
                    0&0&1&0\\
                    0&1&0&0
\end{smallmatrix}\right)$&
$X=\left(\begin{smallmatrix}
            ia&b+in&c+ij&d+ie\\
            -d+ie&f-\frac{i}{2}(a+l)&g+im&ih\\
            -c+ij&k+ip&il&-g+im\\
            -b+in&io&-k+ip&-f-\frac{i}{2}(a+l)
            \end{smallmatrix}\right)$\\ \hline \hline

Defining matrix $E^{(2,2)}$&Representation of $su(2,2)=L_{\J_0
Out_{E^{(2,2)}}}$\\ \hline $E^{(2,2)}_1=\left(\begin{smallmatrix}
                    1&0&0&0\\
                    0&1&0&0\\
                    0&0&-1&0\\
                    0&0&0&-1
\end{smallmatrix}\right)$&
$X=\left(\begin{smallmatrix}
            ia&b+ie&c+ij&d+in\\
            -b+ie&if&g+ik&h+io\\
            c-ij&g-ik&il&m+ip\\
            d-in&h-io&-m+ip&-i(a+f+l)
            \end{smallmatrix}\right)$\\ \hline
$E^{(2,2)}_2=\left(\begin{smallmatrix}
                    1&0&0&0\\
                    0&-1&0&0\\
                    0&0&0&1\\
                    0&0&1&0
\end{smallmatrix}\right)$&
$X=\left(\begin{smallmatrix}
            ia&b+ie&c+in&d+ij\\
            b-ie&if&g+io&h+ik\\
            -d+ij&-h+ik&l-\frac{i}{2}(a+f)&im\\
            -c+in&g-io&ip&-l-\frac{i}{2}(a+f)
            \end{smallmatrix}\right)$\\ \hline
$E^{(2,2)}_3=\left(\begin{smallmatrix}
                    0&1&0&0\\
                    1&0&0&0\\
                    0&0&0&1\\
                    0&0&1&0
\end{smallmatrix}\right)$&
$X=\left(\begin{smallmatrix}
            a+if&ib&c+io&d+ik\\
            ie&-a+if&g+in&h+ij\\
            -h+ij&-d+ik&l-if&im\\
            -g+in&-c+io&ip&-l-if
            \end{smallmatrix}\right)$\\ \hline
$E^{(2,2)}_4=\left(\begin{smallmatrix}
                    1&0&0&0\\
                    0&0&0&-i\\
                    0&0&-1&0\\
                    0&i&0&0
\end{smallmatrix}\right)$&
$X=\left(\begin{smallmatrix}
            ia&b+in&c+ij&d+ie\\
            id+e&f-\frac{i}{2}(a+l)&g+im&h\\
            c-ij&k+ip&il&-ig-m\\
            -ib-n&o&ik+p&-f-\frac{i}{2}(a+l)
            \end{smallmatrix}\right)$\\ \hline
$E^{(2,2)}_5=\left(\begin{smallmatrix}
                    1&0&0&0\\
                    0&-1&0&0\\
                    0&0&0&i\\
                    0&0&-i&0
\end{smallmatrix}\right)$&
$X=\left(\begin{smallmatrix}
            ia&b+ie&c+in&d+ij\\
            b-ie&if&g+io&h+ik\\
            -id-j&ih+k&l-\frac{i}{2}(a+f)&m\\
            ic+n&-ig-o&p&-l-\frac{i}{2}(a+f)
            \end{smallmatrix}\right)$\\ \hline
\end{tabular}
\medskip
\caption{This is the summary of all the various representations of
the five real forms of $sl(4,\C)$ that we need in order to
describe the fine group gradings of these five real forms. For
each of the representations, a generic element $X$ is expressed,
assuming the parameters $a,\ldots,p$ to be real.}
\label{real_forms-sl-representations}
\end{center}
\end{table}

\medskip

The full overview of the fine group gradings of the five real
forms of $sl(4,\C)$ determined by the individual fine gradings
$\Gamma_j$ of $sl(4,\C)$ is contained in Tables
\ref{real_forms-sl-G1}, \ldots, \ref{real_forms-sl-G8}. We proceed
gradually with $\Gamma_1$, $\Gamma_2$, $\ldots$, $\Gamma_8$, and
we describe the basis vectors of the grading subspaces in suitable
representation of the graded real form. We use the same sets of
matrices $X_1$, $\ldots$, $X_{15}$ as already listed in Section
\ref{complex-sl}.

\begin{table}[ht]
\begin{center}
\begin{tabular}{|c||c|c|c|c|c|c|c|c|}
\hline $\Gamma_1$
grading&$L_1$&$L_2$&$L_3$&$L_4$&$L_5$&$L_6$&$L_7$ &$L_8$\\
\cline{2-9}
subspaces&$L_9$&$L_{10}$&$L_{11}$&$L_{12}$&\multicolumn{3}{c|}{$L_{13}$}&\multicolumn{1}{c}{}\\
\hline \hline

real form $sl(4,\R)$&$\R X_1$&$\R X_2$&$\R X_3$&$\R X_4$&$\R
X_5$&$\R X_6$&$\R X_7$&$\R X_8$\\ \cline{2-9} defined by
$F^{(4,\R)}_1$&$\R X_9$&$\R X_{10}$&$\R X_{11}$&$\R
X_{12}$&\multicolumn{3}{c|}{$\R X_{13}+\R X_{14}+\R
X_{15}$}&\multicolumn{1}{c}{}\\ \cline{1-8}
\end{tabular}
\medskip
\caption{The fine group grading $\Gamma_1$ of $sl(4,\C)$
determines a fine group grading $\Gamma^{\J}_1$ for just one real
form of $sl(4,\C)$, namely for $L_{\J}=sl(4,\R)$. The basis
vectors $X_k$ are those listed in Table
\ref{G1}.}\label{real_forms-sl-G1}
\end{center}
\end{table}

\begin{table}[ht]
\begin{center}
\begin{tabular}{|c||c|c|c|c|c|}
\hline $\Gamma_2$ grading&$L_1$&$L_2$&$L_3$&$L_4$&$L_5$\\
\cline{2-6} subspaces&$L_6$&$L_7$&$L_8$&$L_9$&$L_{10}$\\
\cline{2-6} &$L_{11}$&$L_{12}$&$L_{13}$&$L_{14}$&$L_{15}$\\ \hline
\hline

real form $su(3,1)$&$i\R X_1$&$i\R X_2$&$i\R X_3$&$i\R
X_4$&$(1-i)\R X_5$\\ \cline{2-6} defined by $E^{(3,1)}_3$&$\R
X_6$&$(1+i)\R X_7$&$i\R X_8$&$\R X_9$&$i\R X_{10}$\\ \cline{2-6}
&$\R X_{11}$&$i\R X_{12}$&$(1+i)\R X_{13}$&$\R X_{14}$&$(1-i)\R
X_{15}$\\ \hline \hline

real form $su(2,2)$&$i\R X_1$&$i\R X_2$&$i\R X_3$&$(1-i)\R
X_4$&$\R X_5$\\ \cline{2-6} defined by $E^{(2,2)}_4$&$(1+i)\R
X_6$&$i\R X_7$&$\R X_8$&$i\R X_9$&$\R X_{10}$\\ \cline{2-6} &$i\R
X_{11}$&$(1+i)\R X_{12}$&$\R X_{13}$&$(1-i)\R X_{14}$&$i\R
X_{15}$\\ \hline
\end{tabular}
\medskip
\caption{The fine group grading $\Gamma_2$ of $sl(4,\C)$
determines a fine group grading $\Gamma^{\J}_2$ for two real forms
of $sl(4,\C)$, namely for $L_{\J}=su(3,1)$ and $L_{\J}=su(2,2)$.
This result cannot be obtained via Theorem \ref{real_basis}, as
the bases $X_k$ of the grading subspaces in $\Gamma_2$ are not
real matrices - see Table \ref{G2}.}\label{real_forms-sl-G2}
\end{center}
\end{table}

\begin{table}[ht]
\begin{center}
\begin{tabular}{|c||c|c|c|c|c|c|c|c|}
\hline $\Gamma_3$
grading&$L_1$&$L_2$&$L_3$&$L_4$&$L_5$&$L_6$&$L_7$ &$L_8$\\
\cline{2-9}
subspaces&$L_9$&$L_{10}$&$L_{11}$&$L_{12}$&\multicolumn{3}{c|}{$L_{13}$}&\multicolumn{1}{c}{}\\
\hline \hline

real form $sl(4,\R)$&$\R X_1$&$\R X_2$&$\R X_3$&$\R X_4$&$\R
X_5$&$\R X_6$&$\R X_7$&$\R X_8$\\ \cline{2-9} defined by
$F^{(4,\R)}_1$&$\R X_9$&$\R X_{10}$&$\R X_{11}$&$\R
X_{12}$&\multicolumn{3}{c|}{$\R X_{13}+\R X_{14}+\R
X_{15}$}&\multicolumn{1}{c}{}\\ \hline \hline

real form $su(4,0)$&$i\R X_1$&$\R X_2$&$i\R X_3$&$\R X_4$&$i\R
X_5$&$\R X_6$&$i\R X_7$&$\R X_8$\\ \cline{2-9} defined by
$E^{(4,0)}_1$&$i\R X_9$&$\R X_{10}$&$i\R X_{11}$&$\R
X_{12}$&\multicolumn{3}{c|}{$i\R X_{13}+i\R X_{14}+i\R
X_{15}$}&\multicolumn{1}{c}{}\\ \hline \hline

real form $su(3,1)$&$\R X_1$&$i\R X_2$&$\R X_3$&$i\R X_4$&$\R
X_5$&$i\R X_6$&$i\R X_7$&$\R X_8$\\ \cline{2-9} defined by
$E^{(3,1)}_1$&$i\R X_9$&$\R X_{10}$&$i\R X_{11}$&$\R
X_{12}$&\multicolumn{3}{c|}{$i\R X_{13}+i\R X_{14}+i\R
X_{15}$}&\multicolumn{1}{c}{}\\ \hline \hline

real form $su(2,2)$&$\R X_1$&$i\R X_2$&$\R X_3$&$i\R X_4$&$i\R
X_5$&$\R X_6$&$\R X_7$&$i\R X_8$\\ \cline{2-9} defined by
$E^{(2,2)}_1$&$\R X_9$&$i\R X_{10}$&$i\R X_{11}$&$\R
X_{12}$&\multicolumn{3}{c|}{$i\R X_{13}+i\R X_{14}+i\R
X_{15}$}&\multicolumn{1}{c}{}\\ \cline{1-8}

\end{tabular}
\medskip
\caption{Four of the real forms of $sl(4,\C)$ - i.e.
$L_{\J}=sl(4,\R)$, $L_{\J}=su(4,0)$, $L_{\J}=su(3,1)$, and
$L_{\J}=su(2,2)$ - have a fine group grading $\Gamma^{\J}_3$
determined by the fine group grading $\Gamma_3$ of $sl(4,\C)$;
with basis vectors being multiples of $X_k$ defined in Table
\ref{G3} by complex coefficients as listed
here.}\label{real_forms-sl-G3}
\end{center}
\end{table}

\begin{table}[ht]
\begin{center}
\begin{tabular}{|c||c|c|c|c|c|c|c|c|}
\hline $\Gamma_4$
grading&$L_1$&$L_2$&$L_3$&$L_4$&$L_5$&$L_6$&$L_7$ &$L_8$\\
\cline{2-9}
subspaces&$L_9$&$L_{10}$&$L_{11}$&$L_{12}$&$L_{13}$&\multicolumn{2}{c|}{$L_{14}$}&\multicolumn{1}{c}{}\\
\hline \hline

real form $sl(4,\R)$&$\R X_1$&$\R X_2$&$\R X_3$&$\R X_4$&$\R
X_5$&$\R X_6$&$\R X_7$&$\R X_8$\\ \cline{2-9} defined by
$F^{(4,\R)}_1$&$\R X_9$&$\R X_{10}$&$\R X_{11}$&$\R X_{12}$&$\R
X_{13}$&\multicolumn{2}{c|}{$\R X_{14}+\R
X_{15}$}&\multicolumn{1}{c}{}\\ \hline \hline

real form $su(3,1)$&$\R X_1$&$i\R X_2$&$\R X_3$&$i\R X_4$&$\R
X_5$&$i\R X_6$&$\R X_7$&$i\R X_8$\\ \cline{2-9} defined by
$E^{(3,1)}_2$&$\R X_9$&$i\R X_{10}$&$i\R X_{11}$&$i\R X_{12}$&$\R
X_{13}$&\multicolumn{2}{c|}{$i\R X_{14}+i\R
X_{15}$}&\multicolumn{1}{c}{}\\ \hline \hline

real form $su(2,2)$&$\R X_1$&$i\R X_2$&$\R X_3$&$i\R X_4$&$i\R
X_5$&$\R X_6$&$i\R X_7$&$\R X_8$\\ \cline{2-9} defined by
$E^{(2,2)}_2$&$i\R X_9$&$\R X_{10}$&$i\R X_{11}$&$i\R X_{12}$&$\R
X_{13}$&\multicolumn{2}{c|}{$i\R X_{14}+i\R
X_{15}$}&\multicolumn{1}{c}{}\\ \cline{1-8}

\end{tabular}
\medskip
\caption{The fine group grading $\Gamma_4$ of $sl(4,\C)$
determines a fine group grading $\Gamma^{\J}_4$ for three of the
real forms of $sl(4,\C)$: $L_{\J}=sl(4,\R)$, $L_{\J}=su(3,1)$, and
$L_{\J}=su(2,2)$. Their bases can be obtained from matrices $X_k$
defined in Table \ref{G4} via multiplication by complex
coefficients indicated here.}\label{real_forms-sl-G4}
\end{center}
\end{table}

\begin{table}[ht]
\begin{center}
\begin{tabular}{|c||c|c|c|c|c|c|c|c|}
\hline $\Gamma_5$
grading&$L_1$&$L_2$&$L_3$&$L_4$&$L_5$&$L_6$&$L_7$ &$L_8$\\
\cline{2-9}
subspaces&$L_9$&$L_{10}$&$L_{11}$&$L_{12}$&$L_{13}$&\multicolumn{2}{c|}{$L_{14}$}&\multicolumn{1}{c}{}\\
\hline \hline

real form $sl(4,\R)$&$\R X_1$&$\R X_2$&$\R X_3$&$\R X_4$&$\R
X_5$&$\R X_6$&$\R X_7$&$\R X_8$\\ \cline{2-9} defined by
$F^{(4,\R)}_1$&$\R X_9$&$\R X_{10}$&$\R X_{11}$&$\R X_{12}$&$\R
X_{13}$&\multicolumn{2}{c|}{$\R X_{14}+\R
X_{15}$}&\multicolumn{1}{c}{}\\ \hline \hline

real form $su(2,2)$&$\R X_1$&$i\R X_2$&$\R X_3$&$i\R X_4$&$\R
X_5$&$i\R X_6$&$\R X_7$&$i\R X_8$\\ \cline{2-9} defined by
$E^{(2,2)}_3$&$i\R X_9$&$i\R X_{10}$&$i\R X_{11}$&$i\R
X_{12}$&$i\R X_{13}$&\multicolumn{2}{c|}{$\R X_{14}+\R
X_{15}$}&\multicolumn{1}{c}{}\\ \cline{1-8}

\end{tabular}
\medskip
\caption{The fine group grading $\Gamma_5$ of $sl(4,\C)$
determines a fine group grading $\Gamma^{\J}_5$ for the real forms
$L_{\J}=sl(4,\R)$ and $L_{\J}=su(2,2)$ of the complex algebra
$sl(4,\C)$. The matrices $X_k$ listed in Table \ref{G5} are to be
multiplied by complex coefficients from here, in order to obtain
the basis matrices for the fine group grading of the respective
real form.} \label{real_forms-sl-G5}
\end{center}
\end{table}

\begin{table}[ht]
\begin{center}
\begin{tabular}{|c||c|c|c|c|c|c|c|c|}
\hline $\Gamma_6$
grading&$L_1$&$L_2$&$L_3$&$L_4$&$L_5$&$L_6$&$L_7$ &$L_8$\\
\cline{2-9}
subspaces&$L_9$&$L_{10}$&$L_{11}$&$L_{12}$&$L_{13}$&$L_{14}$&$L_{15}$&\multicolumn{1}{c}{}\\
\hline \hline

real form $sl(4,\R)$&$\R X_1$&$\R X_2$&$\R X_3$&$\R X_4$&$\R
X_5$&$\R X_6$&$\R X_7$&$\R X_8$\\ \cline{2-9} defined by
$F^{(4,\R)}_1$&$\R X_9$&$\R X_{10}$&$\R X_{11}$&$\R X_{12}$&$\R
X_{13}$&$\R X_{14}$&$\R X_{15}$&\multicolumn{1}{c}{}\\ \hline
\hline

real form $su^*(4)$&$\R X_1$&$i\R X_2$&$i\R X_3$&$i\R X_4$&$\R
X_5$&$i\R X_6$&$\R X_7$&$\R X_8$\\ \cline{2-9} defined by
$F^{^*(4)}_1$&$i\R X_9$&$\R X_{10}$&$i\R X_{11}$&$\R X_{12}$&$i\R
X_{13}$&$\R X_{14}$&$i\R X_{15}$&\multicolumn{1}{c}{}\\ \hline
\hline

real form $su(2,2)$&$\R X_1$&$\R X_2$&$i\R X_3$&$\R X_4$&$\R
X_5$&$i\R X_6$&$i\R X_7$&$i\R X_8$\\ \cline{2-9} defined by
$E^{(2,2)}_3$&$i\R X_9$&$\R X_{10}$&$i\R X_{11}$&$\R X_{12}$&$i\R
X_{13}$&$i\R X_{14}$&$i\R X_{15}$&\multicolumn{1}{c}{}\\
\cline{1-8}

\end{tabular}
\medskip
\caption{Three real forms $L_{\J}=sl(4,\R)$, $L_{\J}=su^*(4)$, and
$L_{\J}=su(2,2)$ of $sl(4,\C)$ have a fine group grading
$\Gamma^{\J}_6$ determined by the fine group grading $\Gamma_6$ of
$sl(4,\C)$. The basis matrices of these fine group gradings are
complex multiples of the matrices $X_k$ from Table
\ref{G6}.}\label{real_forms-sl-G6}
\end{center}
\end{table}

\begin{table}[ht]
\begin{center}
\begin{tabular}{|c||c|c|c|c|c|c|c|c|}
\hline $\Gamma_7$
grading&$L_1$&$L_2$&$L_3$&$L_4$&$L_5$&$L_6$&$L_7$ &$L_8$\\
\cline{2-9}
subspaces&$L_9$&$L_{10}$&$L_{11}$&$L_{12}$&$L_{13}$&$L_{14}$&$L_{15}$&\multicolumn{1}{c}{}\\
\hline \hline

real form $sl(4,\R)$&$\R X_1$&$\R X_2$&$\R X_3$&$\R X_4$&$\R
X_5$&$\R X_6$&$\R X_7$&$\R X_8$\\ \cline{2-9} defined by
$F^{(4,\R)}_1$&$\R X_9$&$\R X_{10}$&$\R X_{11}$&$\R X_{12}$&$\R
X_{13}$&$\R X_{14}$&$\R X_{15}$&\multicolumn{1}{c}{}\\ \hline
\hline

real form $su^*(4)$&$i\R X_1$&$\R X_2$&$i\R X_3$&$\R X_4$&$i\R
X_5$&$\R X_6$&$i\R X_7$&$\R X_8$\\ \cline{2-9} defined by
$F^{^*(4)}_1$&$\R X_9$&$i\R X_{10}$&$\R X_{11}$&$i\R X_{12}$&$i\R
X_{13}$&$\R X_{14}$&$i\R X_{15}$&\multicolumn{1}{c}{}\\ \hline
\hline

real form $su(4,0)$&$i\R X_1$&$i\R X_2$&$\R X_3$&$\R X_4$&$i\R
X_5$&$\R X_6$&$i\R X_7$&$\R X_8$\\ \cline{2-9} defined by
$E^{(4,0)}_1$&$i\R X_9$&$\R X_{10}$&$\R X_{11}$&$i\R X_{12}$&$i\R
X_{13}$&$i\R X_{14}$&$i\R X_{15}$&\multicolumn{1}{c}{}\\ \hline
\hline

real form $su(2,2)$&$\R X_1$&$\R X_2$&$i\R X_3$&$i\R X_4$&$\R
X_5$&$i\R X_6$&$\R X_7$&$i\R X_8$\\ \cline{2-9} defined by
$E^{(2,2)}_1$&$i\R X_9$&$\R X_{10}$&$\R X_{11}$&$i\R X_{12}$&$i\R
X_{13}$&$i\R X_{14}$&$i\R X_{15}$&\multicolumn{1}{c}{}\\
\cline{1-8}

\end{tabular}
\medskip
\caption{The fine group grading $\Gamma_7$ of $sl(4,\C)$
determines a fine group grading $\Gamma^{\J}_7$ for four of the
real forms of $sl(4,\C)$, namely $L_{\J}=sl(4,\R)$,
$L_{\J}=su^*(4)$, $L_{\J}=su(4,0)$, and $L_{\J}=su(2,2)$. Their
bases are given by means of the matrices $X_k$ from Table
\ref{G7}.}\label{real_forms-sl-G7}
\end{center}
\end{table}

\begin{table}[ht]
\begin{center}
\begin{tabular}{|c||c|c|c|c|c|}
\hline $\Gamma_8$ grading&$L_1$&$L_2$&$L_3$&$L_4$&$L_5$\\
\cline{2-6} subspaces&$L_6$&$L_7$&$L_8$&$L_9$&$L_{10}$\\
\cline{2-6} &$L_{11}$&$L_{12}$&$L_{13}$& \multicolumn{2}{c|}{$L_{14}$}\\
\hline \hline

real form $sl(4,\R)$&$\R X_1$&$\R X_2$&$\R X_3$&$\R X_4$&$\R
X_5$\\ \cline{2-6} defined by $F^{(4,\R)}_1$&$\R X_6$&$\R X_7$&$\R
X_8$&$\R X_9$&$\R X_{10}$\\ \cline{2-6} &$\R X_{11}$&$\R
X_{12}$&$\R X_{13}$&\multicolumn{2}{c|}{$\R X_{14}+\R X_{15}$}\\
\hline \hline

real form $su^*(4)$&$(1-i)\R X_1$&$(1+i)\R X_2$&$(1+i)\R
X_3$&$(1-i)\R X_4$&$(1+i)\R X_5$\\ \cline{2-6} defined by
$F^{^*(4)}_2$&$(1-i)\R X_6$&$(1-i)\R X_7$&$(1+i)\R X_8$&$\R
X_9$&$i\R X_{10}$\\ \cline{2-6} &$i\R X_{11}$&$\R X_{12}$&$\R
X_{13}$&\multicolumn{2}{c|}{$i\R X_{14}+i\R X_{15}$}\\ \hline
\hline

real form $su(3,1)$&$i\R X_1$&$i\R X_2$&$\R X_3$&$\R X_4$&$i\R
X_5$\\ \cline{2-6} defined by $E^{(3,1)}_2$&$\R X_6$&$\R X_7$&$\R
X_8$&$i\R X_9$&$i\R X_{10}$\\ \cline{2-6} &$\R X_{11}$&$\R
X_{12}$&$i\R X_{13}$&\multicolumn{2}{c|}{$i\R X_{14}+i\R X_{15}$}\\
\hline \hline

real form $su(2,2)$&$(1+i)\R X_1$&$(1+i)\R X_2$&$(1-i)\R
X_3$&$(1-i)\R X_4$&$(1-i)\R X_5$\\ \cline{2-6} defined by
$E^{(2,2)}_5$&$(1+i)\R X_6$&$(1-i)\R X_7$&$(1+i)\R X_8$&$i\R
X_9$&$i\R X_{10}$\\ \cline{2-6} &$\R X_{11}$&$i\R X_{12}$&$\R
X_{13}$&\multicolumn{2}{c|}{$\R X_{14}+\R X_{15}$}\\ \hline

\end{tabular}
\medskip
\caption{Also the fine group grading $\Gamma_8$ of $sl(4,\C)$
determines a fine group grading $\Gamma^{\J}_8$ for four of the
real forms of $sl(4,\C)$, this time for $L_{\J}=sl(4,\R)$,
$L_{\J}=su^*(4)$, $L_{\J}=su(3,1)$, and $L_{\J}=su(2,2)$. Their
bases derive from the matrices $X_k$ given in Table \ref{G8} via
multiplication by complex coefficients listed
here.}\label{real_forms-sl-G8}
\end{center}
\end{table}

\medskip

Comparing these results with the ones obtained by the 'MAD-group'
method (exhausting the whole list of MAD-groups $\G_1^{\R}$,
$\G_3^{\R}$, $\G_4^{\R}$, $\G_5^{\R}$, $\G_6^{\R}$, and $\G_7^{\R}$
acting on one or more of the real forms), we see that the 'real
basis' method and the more general 'fundamental' method bring in six
additional fine group gradings of the real forms, namely those
determined by the gradings $\Gamma_2$ and $\Gamma_8$ of $sl(4,\C)$.
The fact that for the complex MAD-groups $\G_2$ and $\G_8$ there
exist no MAD-groups on real forms can be deduced from the universal
grading groups $\Z_4^2$ and $\Z_4 \times \Z_2^2$ corresponding to
the gradings $\Gamma_2$ and $\Gamma_8$ respectively, since the
fourth root of unity does not belong to $\R$.\\

Thereby we show that for the real forms of the simple complex Lie
algebra $sl(4,\C)$ there exist fine group gradings which are not
generated by MAD-groups of the real forms.

\begin{table}[ht]
\begin{center}
\begin{tabular}{|l|l|l|}
\hline real form &fine gradings obtained by&fine gradings obtained
by 'real\\of $sl(4,\C)$&'MAD-group' method&basis' or 'fundamental'
method only\\ \hline \hline
$sl(4,\R)$&$\Gamma_1^{\J},\Gamma_3^{\J},\Gamma_4^{\J},\Gamma_5^{\J},
\Gamma_6^{\J},\Gamma_7^{\J}$&$\Gamma_8^{\J}$\\ \hline

$su^{*}(4)$&$\Gamma_6^{\J},\Gamma_7^{\J}$&$\Gamma_8^{\J}$\\ \hline

$su(4,0)$&$\Gamma_3^{\J},\Gamma_7^{\J}$&\\ \hline

$su(3,1)$&$\Gamma_3^{\J},\Gamma_4^{\J}$&$\Gamma_2^{\J},\Gamma_8^{\J}$\\
\hline

$su(2,2)$&$\Gamma_3^{\J},\Gamma_4^{\J},\Gamma_5^{\J},\Gamma_6^{\J},
\Gamma_7^{\J}$&$\Gamma_2^{\J},\Gamma_8^{\J}$\\ \hline
\end{tabular}
\medskip
\caption{Summary of the fine group gradings of real forms of
$sl(4,\C)$: Those determined by $\Gamma_1$, $\Gamma_3$, $\Gamma_4$,
$\Gamma_5$, $\Gamma_6$, $\Gamma_7$ can be obtained as decompositions
of the real forms into simultaneous eigensubspaces of automorphisms
from MAD-groups of the real forms, which is not the case of the fine
group gradings determined by $\Gamma_2$ and
$\Gamma_8$.}\label{real_forms-sl-2methods}
\end{center}
\end{table}

\subsection{Fine Group Gradings of the Real Forms of $sp(4,\C)$ and $o(4,\C)$}
\label{real-sp+o}

Just like the complex Lie algebras $L=sp_K(n,\C)$ and
$L=o_K(n,\C)$ are subalgebras of the complex Lie algebra
$sl(n,\C)$, also the real forms of $L$ are subalgebras of the real
forms of $sl(n,\C)$. See Table \ref{real_forms-sl+sp+o} for the
full overview of real forms of $L=sp_K(4,\C)$ and $L=o_K(4,\C)$
and their 'source' real forms of $sl(4,\C)$.

\begin{table}[ht]
\begin{center}
\begin{tabular}{|c||c|c|}
\hline real form of $sl(4,\C)$&real form of $sp(4,\C)$&real form
of $o(4,\C)$\\ \hline \hline $sl(4,\R)$&$sp(4,\R)$&\\ \hline
$su^*(4)$&&$so^*(4)$\\ \hline $su(4,0)$&$usp(4,0)$&$so(4,0)$\\
\hline $su(3,1)$&&$so(3,1)$\\ \hline
$su(2,2)$&$usp(2,2)$&$so(2,2)$\\ \hline
\end{tabular}
\medskip
\caption{The full list of the real forms of $sp(4,\C)$ and of
$o(4,\C)$, and their relation to the real forms of
$sl(4,\C)$.}\label{real_forms-sl+sp+o}
\end{center}
\end{table}

Having this direct link between the real forms of $sl(4,\C)$ and
those of its subalgebras $L\subset sl(4,\C)$, we easily obtain
fine gradings of the real forms of $L$, because they are displayed
by the fine gradings of the respective 'source' real forms of
$sl(4,\C)$.

The choice of the representation matrix $K$ is given already by
the way the complex subalgebra is displayed by the fine gradings
of the complex $sl(4,\C)$ - see $K_0=J$ for $sp(4,\C)$ in Table
\ref{complex-sp} and $K_1$, $K_2$, $K_3$ for $o(4,\C)$ in Table
\ref{complex-o}.

\subsubsection{Fine Group Gradings of the Real Forms of $sp(4,\C)$} \label{real-sp}

When describing the fine group gradings of the real forms of
$sp(4,\C)$, we stay in just one representation
$sp_{K_0}(4,\C)=sp_J(4,\C)$ - as corresponds to the complex case.
As the complex subalgebra $sp(4,\C)$ of $sl(4,\C)$ is displayed
just by three fine group gradings of $sl(4,\C)$ - namely
$\Gamma_5$, $\Gamma_6$, $\Gamma_7$, it is only these three cases
that come into play also for the real forms:

\begin{itemize}
    \item The real form $sp(4,\R)$ has fine group gradings
    determined by the original fine group gradings $\Gamma_5$,
    $\Gamma_6$, $\Gamma_7$ of $sl(4,\C)$:\\
    $\Gamma_5, \Gamma_6, \Gamma_7\quad\rightarrow\quad$ $sp(4,\R)=sp_J(4,\C)\cap
    sl_{F_1}(4,\R)$
    \item The real form $usp(4,0)$ has a fine group grading
    determined by $\Gamma_7$:\\
    $\Gamma_7\qquad\qquad\rightarrow\quad$ $usp(4,0)=sp_J(4,\C)\cap
    su_{E_1}(4,0)$
    \item The real form $usp(2,2)$ has fine group gradings
    determined by $\Gamma_5$, $\Gamma_6$, $\Gamma_7$:\\
    $\Gamma_5, \Gamma_6\quad\quad\ \rightarrow\quad$ $usp(2,2)=sp_J(4,\C)\cap
    su_{E_3}(2,2)$\\
    $\Gamma_7\qquad\qquad\rightarrow\quad$ $usp(2,2)=sp_J(4,\C)\cap
    su_{E_1}(2,2)$
\end{itemize}

\begin{table}[ht]
\begin{center}
\begin{tabular}{|l|l|}
\hline real form of $sp(4,\C)$ &fine gradings displayed by fine
gradings of real forms of $sl(4,\C)$\\ \hline \hline
$sp(4,\R)$&$\Gamma_5^{\J}, \Gamma_6^{\J},\Gamma_7^{\J}$\\ \hline

$usp(4,0)$&$\Gamma_7^{\J}$\\ \hline

$usp(2,2)$&$\Gamma_5^{\J}, \Gamma_6^{\J}, \Gamma_7^{\J}$\\ \hline
\end{tabular}
\medskip
\caption{Summary of fine group gradings of the real forms of
$sp(4,\C)$ displayed by the fine group gradings of the real forms
of $sl(4,\C)$.}\label{real_forms-sp}
\end{center}
\end{table}

\subsubsection{Fine Group Gradings of the Real Forms of $o(4,\C)$} \label{real-o}

Here we must work with more representations of the algebra
$o(4,\C)$ already - like in the complex case. The different
representations that come into play are those defined by symmetric
matrices $K_1$, $K_2$, and $K_3$ listed in Table \ref{complex-o}.
The complex subalgebra $o(4,\C)$ is displayed by six fine group
gradings of $sl(4,\C)$ - namely $\Gamma_3$, $\Gamma_4$,
$\Gamma_5$, $\Gamma_6$, $\Gamma_7$, $\Gamma_8$. Thus just these
six group gradings display fine group gradings for real forms of
$o(4,\C)$, in those cases that are relevant for the respective
real form.

\begin{itemize}
    \item The real form $so^*(4)$ has fine group gradings
    determined by the original fine group gradings $\Gamma_6$,
    $\Gamma_7$, $\Gamma_8$ of $sl(4,\C)$:\\
    $\Gamma_6\qquad\qquad\rightarrow\quad$ $so^*(4)=o_{K_3}(4,\C)\cap
    su^*_{F_1}(4)$\\
    $\Gamma_7\qquad\qquad\rightarrow\quad$ $so^*(4)=o_{K_1}(4,\C)\cap
    su^*_{F_1}(4)$\\
    $\Gamma_8\qquad\qquad\rightarrow\quad$ $so^*(4)=o_{K_2}(4,\C)\cap
    su^*_{F_2}(4)$
    \item The real form $so(4,0)$ has fine group gradings
    determined by $\Gamma_3$ and $\Gamma_7$:\\
    $\Gamma_3, \Gamma_7\quad\quad\ \rightarrow\quad$ $so(4,0)=o_{K_1}(4,\C)\cap
    su_{E_1}(4,0)$
    \item The real form $so(3,1)$ has fine group gradings
    determined by $\Gamma_3$, $\Gamma_4$, $\Gamma_8$:\\
    $\Gamma_3\qquad\qquad\rightarrow\quad$ $so(3,1)=o_{K_1}(4,\C)\cap
    su_{E_1}(3,1)$\\
    $\Gamma_4, \Gamma_8\quad\quad\ \rightarrow\quad$ $so(3,1)=o_{K_2}(4,\C)\cap
    su_{E_2}(3,1)$
    \item The real form $so(2,2)$ has fine group gradings
    determined by $\Gamma_3$, $\Gamma_4$, $\Gamma_5$, $\Gamma_6$, $\Gamma_7$, and $\Gamma_8$:\\
    $\Gamma_3, \Gamma_7\quad\quad\ \rightarrow\quad$ $so(2,2)=o_{K_1}(4,\C)\cap
    su_{E_1}(2,2)$\\
    $\Gamma_4\qquad\qquad\rightarrow\quad$ $so(2,2)=o_{K_2}(4,\C)\cap
    su_{E_2}(2,2)$\\
    $\Gamma_5, \Gamma_6\quad\quad\ \rightarrow\quad$ $so(2,2)=o_{K_3}(4,\C)\cap
    su_{E_3}(2,2)$\\
    $\Gamma_8\qquad\qquad\rightarrow\quad$ $so(2,2)=o_{K_2}(4,\C)\cap
    su_{E_5}(2,2)$
\end{itemize}

\begin{table}[ht]
\begin{center}
\begin{tabular}{|l|l|}
\hline real form of $o(4,\C)$ &fine gradings displayed by fine
gradings of real forms of $sl(4,\C)$\\ \hline \hline
$so^*(4)$&$\Gamma_6^{\J}, \Gamma_7^{\J},\Gamma_8^{\J}$\\ \hline

$so(4,0)$&$\Gamma_3^{\J}, \Gamma_7^{\J}$\\ \hline

$so(3,1)$&$\Gamma_3^{\J}, \Gamma_4^{\J}, \Gamma_8^{\J}$\\ \hline

$so(2,2)$&$\Gamma_3^{\J}, \Gamma_4^{\J}, \Gamma_5^{\J},
\Gamma_6^{\J}, \Gamma_7^{\J}, \Gamma_8^{\J}$\\ \hline
\end{tabular}
\medskip
\caption{Summary of fine group gradings of the real forms of
$o(4,\C)$ displayed by the fine group gradings of the real forms
of $sl(4,\C)$.}\label{real_forms-o}
\end{center}
\end{table}

\section{Concluding  remarks}\label{concl}


In the whole article, we deal only with fine group gradings of
real Lie algebras, but we do not study non-group gradings. Whereas
on complex Lie algebras there is a one-to-one correspondence
between fine group gradings and MAD-groups, on real forms of
$sl(4,\C)$ we have found fine group gradings which are not
generated by any MAD-group on the respective real form. The
question whether our list of fine group gradings of these real
forms is complete still remains open. This problem requires
further investigation, similarly as the question of existence of
non-group fine gradings.

Even for the known gradings, a number of additional questions can
be raised. Let us point out a few:

\begin{itemize}

\item The gradings of Lie algebras should be extended to their
representations, finite as well as infinite dimensional ones.

\item Lie algebras often need to be extended (semidirect product)
by an Abelian algebra (translations). Best known case is the
Poincar\'e Lie algebra of space-time transformations extending
$o(3,1)$. Fine gradings of such algebras would be of interest to
know.

\item Fine gradings which decompose the Lie algebra into
one-dimensional spaces define a basis of the algebra reflecting
its unique structure, certainly a particular simplicity of
corresponding structure constants.

\item Gradings of an algebra $L$ frequently display particular
subalgebras of $L$. That is, such subalgebras are formed by one or
several grading subspaces. All of the fine gradings of $L$
together display many subalgebras. Which maximal subalgebras are
displayed, and which are missing?

\item  Casimir operators should have rather different form in
different fine gradings. Casimir operators of displayed
subalgebras should be clearly recognized there.

\end{itemize}
\bigskip

\section*{Acknowledgements}

We are grateful to the referee, who drew our attention to the
problems connected with the difference between the terms 'grading'
and 'group grading'. His / her comments have significantly
improved the quality of our text. The referee has, among other,
pointed out that the Theorem \ref{real_basis} can be stated as
equivalence, and has provided a proof of the implication that was
originally missing.

The authors acknowledge financial support of the Doppler Institute
by the grant LC06002 of the Ministry of Education, Youth and
Sports of the Czech Republic.


\end{document}